\documentclass[aps,prl,twocolumn,superscriptaddress,longbibliographys]{revtex4-2}

\usepackage{graphicx} 
\usepackage[utf8]{inputenc}
\usepackage{bm}
\usepackage{natbib}
\usepackage{color}
\usepackage{epstopdf}
\usepackage{amsmath}
\usepackage{amssymb}
\usepackage{bm}
\usepackage{enumerate}
\usepackage{chngpage}
\usepackage{url}
\usepackage{braket}
\usepackage{filecontents}
\definecolor{darkGreen}{RGB}{0,110,0}
\definecolor{darkBlue}{RGB}{0,0,130}
\usepackage[colorlinks=true, urlcolor=darkBlue,citecolor=darkGreen,linkcolor=darkBlue]{hyperref}

\newcommand{\bk}{{\bm{k}}}

\begin{document}

\title{Upper critical field and pairing symmetry of Ising superconductors}

\author{Lena Engstr\"om}
\email{lena.engstrom@universite-paris-saclay.fr}
\affiliation {Université Paris-Saclay, CNRS, Laboratoire de Physique des Solides, 91405 Orsay, France}
\author{Ludovica Zullo}
\affiliation{
Institut für Theoretische Physik und Astrophysik and Würzburg-Dresden Cluster of Excellence ct.qmat,
Universität Würzburg, 97074 Würzburg, Germany}
\author{Tristan Cren}
\affiliation{Sorbonne Université, CNRS, Institut des Nanosciences de Paris, UMR7588, F-75252 Paris, France}
\author{Andrej Mesaros}
\affiliation{Université Paris-Saclay, CNRS, Laboratoire de Physique des Solides, 91405 Orsay, France}
\author{Pascal Simon}
\affiliation{Université Paris-Saclay, CNRS, Laboratoire de Physique des Solides, 91405 Orsay, France}

\begin{abstract}
Motivated by the fact that the measured critical field $H_{c2}$ in various transition metal dichalcogenide (TMD) superconductors is poorly understood, we reexamine its scaling behavior with temperature and spin-orbit coupling (SOC). By computing the spin-susceptibility in a multipocket system, we find that segments of the Fermi Surface (FS) at which the SOC has nodal points can have a contribution orders of magnitude larger than the remaining FS, hence setting the $H_{c2}$, assuming the presence of a conventional singlet superconducting order parameter. Nodal lines of an Ising SOC in the Brillouin zone are imposed by symmetry, so they cause such nodal points whenever they intersect an FS pocket, which is indeed the case in monolayer NbSe$_2$ and TaS$_2$, but not in gated MoS$_2$ and WS$_2$. Our analysis reinterprets existing measurements, concluding that a dominant singlet-order parameter on pockets with SOC nodes is consistent with the $H_{c2}(T)$ data for all monolayer Ising superconductors, in contrast to previous contradictory pairing assumptions. Finally, we predict a doping-dependent experimental signature of our theory.
\end{abstract}

\maketitle
\begin{figure*}
    \includegraphics[width=\linewidth]{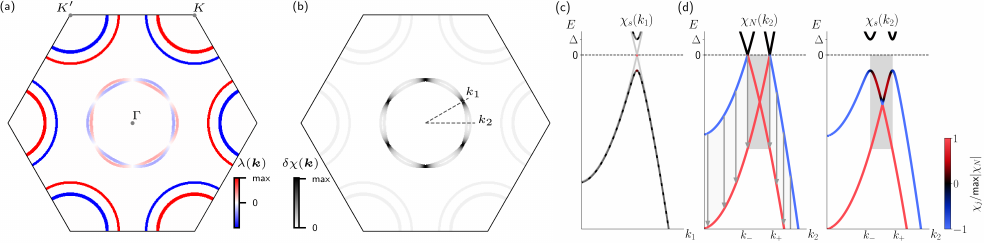}
    \caption{\label{fig:FSspinChi}a) FS of a monolayer Ising superconductor, of the same type as NbSe$_2$. Two fully spin-polarized bands $\xi_\pm = \xi_k \pm \lambda (\boldsymbol{k})$ are split by the Ising SOC $\lambda (\boldsymbol{k})$. We model $\lambda^\Gamma (\theta) = \lambda^\Gamma_0 \cos \left(3 \theta \right)$ and $\lambda^K (\theta) = \lambda^K_0 $, where $\lambda^\Gamma_0 \approx \lambda^K_0 /3$. b) The contribution to the difference in susceptibility between a uniform singlet superconducting state and the normal state. At the nodal points of the SOC $\lambda^\Gamma (\theta)=0$ the superconductor is a conventional singlet order. As the SOC is large everywhere else, $\delta \chi (T)$, and consequently the critical field $H_{c2}$, is determined almost entirely by the 6 nodal points. c) \& d) The normal and superconducting state susceptibilities $\chi_N (E), \chi_S (E)$ are shown on the $\Gamma$-pocket along radial lines, passing through a nodal point (c), or between nodal points (d). For spin-split bands in (d), the energy range where the combined susceptibility of the bands is non-zero extends to $\xi_{k_\pm}=\pm\lambda_0$ (shaded). In (c) the plots are individually scaled as max$|\chi_N (E)| \gg$max$|\chi_S (E)|$.}
\end{figure*}

\textit{Introduction.} 
For a given temperature, the upper critical field $H_{c2}$ refers to the maximum magnetic field  below which a  material remains superconducting. For a singlet superconductor with spin-degenerate bands $H_{c2} (T=0)$ is determined by the Pauli limit $H_p = 1.76 k_{\text{B}} T_c / \sqrt{2} \mu_B$ \cite{Chandrasekhar1962, Clogston1962}. However, it was long recognized that this Pauli limit can be exceeded in presence of the spin-orbit coupling (SOC)\cite{Klemm1975,Frigeri2004, Frigeri2004a}.
Many recent experiments on two-dimensional (2D) layered superconductors have measured a $H_{c2}> H_p$. This is particularly noticeable in monolayer superconducting transition metal dichalcogenides (TMDs), also known as Ising superconductors, showing very high in-plane critical fields\cite{Xi2016, Yang2018, DeLaBarrera2018}. These materials are characterized by a large SOC originating from the broken inversion symmetry. Though remarkable, in these systems it remains difficult to extract significant information about the nature of the superconducting order parameter, such as being singlet or triplet \cite{Sigrist2014,Mockli2020}.

Additional puzzling issues arise in the 1H-MX$_2$ compounds, M$=$ Nb, Ta, Mo, W, and X$=$ S, Se. For the few-layer stacking configurations, an in-plane magnetic field has a negligible coupling to the orbital degrees of freedom of electrons and instead couples only to spin\cite{Matsuoka2020}. The measured $H_{c2}$ for monolayer NbSe$_2$ and TaS$_2$ indicates a square root scaling with the SOC while theories for a spin-split Fermi surface (FS) predict a linear scaling \cite{Frigeri2004, Ilic2017,Mockli2019}. Moreover, if the bands are spin-split, the intervalley scattering  required to fit the experimental data should be  several orders of magnitude higher in NbSe$_2$ than in WS$_2$\cite{DeLaBarrera2018,Ilic2017}, something for which there has been no physical explanation.
\begin{table}[!h]
\begin{tabular}{c|ccc}
        $\delta \chi \propto$              & no SOC                   & constant SOC                      & nodal SOC                \\
                     \hline
singlet              &  $\Delta^0$          & $\Delta^2$ & $\Delta$   \\
triplet, $\bm{h} \perp \bm{d}$ & $\Delta^2$ & $\Delta^2$ & $\Delta^2$ \vspace{0.05\linewidth}
\end{tabular}
\begin{tabular}{c|ccc}
      $H_{c2} /H_p \propto$               & no SOC                   & constant SOC                      & nodal SOC               \\
                     \hline
singlet              & 1          & $\frac{\lambda}{\Delta}$ & $\sqrt{\frac{\lambda}{\Delta}}$   \\
triplet, $\bm{h} \perp \bm{d}$ & $\frac{\epsilon}{\Delta}$ & $\frac{\sqrt{\epsilon^2 - \lambda^2}}{\Delta}$ & $\frac{\sqrt{\epsilon^2 - \lambda^2}}{\Delta}$
\end{tabular}
\caption{\label{tab:scalingSOC}The leading order term of the expansion of the susceptibility difference $\delta \chi$ in a small superconducting order parameter $\Delta$, for a magnetic field $\bm{h} \perp \bm{g}$ with $\bm{g}$ the spin-direction of the SOC (top). The resulting scaling with SOC strength $\lambda$ of the critical field, assuming a single pocket of the FS, with a given profile of SOC (absent; constant over the pocket; having nodal points). For a multi-pocket FS one pocket can have a $\delta \chi$ orders of magnitude larger than for the other pockets. Of note are pockets with a nodal SOC, where singlet pairing produces a unique scaling.
}
\end{table}

In this Letter, we calculate the critical field via the susceptibility taking into account the contribution of all pockets present at the Fermi surface. We find that the $\Gamma$-pocket, when present, almost entirely dominates the susceptibility and hence determines the critical field $H_{c2}(T)$ in Ising superconductors. This is due to the fact that by symmetry the Ising SOC must have nodal points (at which there is no spin splitting) on the $\Gamma$-pocket\footnote{Note that a Rashba SOC opens a small gap  at these nodes, but our conclusion remains for a dominant Ising SOC, which is the case in TMDs. See \cite{SupMat} for details and references \protect\cite{Huertas-Hernando2006,Samokhin2009} therein.}. Despite the $H_{c2}$ having been calculated decades ago for an isolated pocket with SOC nodes\cite{Bulaevskil1977}, its potential importance, nor the interplay with other pockets, has not been fully appreciated for superconducting TMDs\cite{Mockli2020}. For example, due to the position of its Fermi level, NbSe$_2$ has the additional pocket around the $\Gamma$-point while WS$_2$ does not. For a singlet pairing, the susceptibility at these nodal points is simply that of a conventional superconductor which therefore suppresses $H_{c2}$. Our results on the scaling of the susceptibility for singlet and triplet order parameters are summarized in Table~\ref{tab:scalingSOC}. In general, Ising superconductors are likely to host mixed-parity orders, where the singlet-triplet mixing depends on the SOC and the interpocket coupling\cite{Wickramaratne2020}. In this work, we consider single-component orders, hence focusing on the leading component. We find that the scaling of the critical field $H_{c2}$ with SOC indicates that the effective pairing interaction between pockets is not negligible, and spin-singlet is the dominant order parameter in each compound in the family of monolayer Ising superconductors. We get values that are consistent with existing experimental data over all monolayer compounds. Finally, we predict that our scaling with SOC has a direct consequence observable under doping.

\textit{Upper critical field from susceptibility.} The critical field $H_{c2}$ can be defined as the field at which the thermodynamic potentials $\Omega_i(T,H)$, for the normal state $(i=N)$ and for a proposed superconducting state $(i=S)$, are equal in value\cite{Clogston1962,Chandrasekhar1962}. Throughout, we will be assuming a temperature near the critical temperature $T_c$, so that $H_{c2}$ is small enough to expand the potential in terms of the susceptibility. This standard procedure gives
\begin{equation}
    H_{c2} (T) = \sqrt{\frac{\Omega_0(T) }{\delta \chi(T) }},
\end{equation}
where $\Omega_0(T)$ is the total condensation energy\cite{SupMat,Ortega2020}, and $\delta \chi(T) = \chi_N (T)- \chi_S(T)$ is the difference in susceptibility between the two states. Near $T_c$, it is insightful to further expand the susceptibility difference in orders of the superconducting order parameter $\Delta$:
\begin{equation}\label{eq:chiOrd}
    \delta \chi(T) =   \delta \chi_{0}(T)  +  \delta \chi_{1}(T) + \delta \chi_{2}(T) + \mathcal{O} \left( \Delta^3\right),
\end{equation}
where $\delta \chi_{m}=\mathcal{O}(\Delta^m)$. The leading term then determines the value of the critical field, for example, if it is the zero-th order then $H_{c2} (T) =\sqrt{ \Omega_0 (T )/ \delta \chi_{0}(T) } \approx H_p$, while if it is the first order then $\sqrt{\Omega_0 (T )/ \delta \chi_{1}(T) } \gg H_p$. This connection is summarized for known cases in Table~\ref{tab:scalingSOC}, which includes various scenarios for the SOC, which we discuss now.

\textit{SOC and multiple FS pockets in TMDs.} We now consider the effective TMD transition metal bandstructure throughout the Brillouin zone (BZ), $\xi_{\bk,\zeta} = \xi_\bk + \zeta \lambda(\bm{k})$, with $\zeta=\pm$ labeling the bands that are split by the SOC $\lambda(\bm{k})$. Note a key property: the Ising SOC is described by the scalar function $\lambda(\bm{k})$ that must be odd in $\bm{k}$, and hence must have nodal lines emanating from the $\Gamma$ point in the Brillouin zone. 

Furthermore, for given bands and doping level, the FS can form multiple pockets, typically in TMDs around the $\Gamma$ and $K$ points in the BZ (see Fig.~\ref{fig:FSspinChi}a). For each pocket, labeled $j=\Gamma,K\ldots$, we consider a parabolic band $\xi^j_k$\footnote{The calculations can be compared to a realistic bandstructure resulting only in small corrections to the critical field\protect\cite{Rahn2012,SupMat}.} in the energy range $\pm \epsilon$ which captures the SOC splitting, namely, $\epsilon > \lambda^j_0$\footnote{The energy range must reach some physically relevant cut-off energy $\epsilon = \lambda_0^j + \epsilon_c$. In conventional superconductors with $\lambda_0=0$, the Debye frequency $\epsilon_c = \omega_{\text{D}}$ is chosen.}, where the SOC function at the pocket simplifies to $\lambda^j(\bm{k})=\lambda^j_0 g^j(\theta)$, with $\theta$ the polar angle around the pocket, $\lambda_0^j$ is the maximal amplitude of the SOC at the pocket, and $g^j$ a dimensionless angular function.

The normal state susceptibility is a sum of contributions from each pocket, $\chi_N=\sum_j \chi_N^j$. We consider only in-plane magnetic fields $h$, so due to the two spin-split bands $\zeta=\pm$ at a pocket $j$, the susceptibility can further be usefully divided into two contributions of different physical origin. We have the intraband (Pauli) and the interband (van Vleck) contributions\cite{Sigrist2009} $\chi^j_N = \chi_N^{j,\text{intra}}  + \chi_N^{j,\text{inter}}$, with
\begin{align}
    \chi_N^{j,\text{intra}} = &\sum_{\zeta, \bm{k}} \left( \left. \frac{\partial \xi^j_{\zeta,h}}{\partial h} \right|_{h=0} \right)^2 \frac{\partial f(\xi^j_{\zeta})}{\partial \xi^j_{\zeta}}~,\\
    \chi_N^{j,\text{inter}} = &\sum_{\zeta, \bm{k}} \left. \frac{\partial^2 \xi^j_{\zeta,h}}{\partial h^2} \right|_{h=0}  f(\xi^j_{\zeta}),
\end{align}
where $f$ the Fermi-Dirac distribution. The intraband term is strongly peaked at the FS, however due to the in-plane field any finite splitting $\lambda^j(\bm{k})\neq 0$ makes this term vanish. In contrast, the interband term has {\it a priori} contributions from all filled states in the pocket. However, we find cancellations everywhere except in the energy region containing the band splitting (see Fig.~\ref{fig:FSspinChi}d). Physically, the finite splitting makes the susceptibility interband, originating from the canting of spins within the energy range of the splitting.

\textit{Susceptibility in the superconductor.} Throughout we will consider only pairing between $\bm{k}$ and $-\bm{k}$ electrons, hence we can define the pairing functions $\Delta^j$, for example, for the $j=\Gamma$ pocket, and $j=K$ for pairing between $K$ and $-K$ pockets. We will assume that the temperature dependence of the pairing functions $\Delta^j$ follows the mean-field solution for $h=0$ with $\Delta^j(T=0) = \Delta^j_0 = 1.76 k_{\text{B}} T^j_c$. We will discuss in detail the relationships between different $\Delta^j$ in the next section.

Focusing on one pairing $\Delta^j$, the condensation energy is actually independent of the SOC $\lambda^j_0$\cite{SupMat}. Thus, the condensation energy $\Omega^j_0 (T)$ is simply proportional to the density of states of the $j$ pocket and to $(\Delta^j_0)^2$ (see \cite{SupMat}).

The susceptibility in the superconducting state also has two contributions for each fixed $j$, $\chi^j_S = \chi_S^{j,\text{intra}}  + \chi_S^{j,\text{inter}}$, defined for the BdG bands $E^j_{\zeta}$ as\cite{Samokhin2021}:
\begin{equation}
    \chi_S^{j,\text{intra}} = \sum_{\zeta, \bm{k}} \left( \left. \frac{\partial E^j_{\zeta,h}}{\partial h} \right|_{h=0} \right)^2 \frac{\partial f(E^j_{\zeta})}{\partial E^j_{\zeta}},
\end{equation}
\begin{align}\label{eq:chiSinter}
    \chi_S^{j,\text{inter}} =& \sum_{\zeta, \bm{k}} \left( \left. \frac{\partial^2 E^j_{\zeta,h}}{\partial h^2} \right|_{h=0}  f(E^j_{\zeta}) \right. \\ \notag
     & \left.  + \frac{1}{2}\left( \left. \frac{\partial^2 \xi^j_{\zeta,h}}{\partial h^2} \right|_{h=0} - \left. \frac{\partial^2 E^j_{\zeta,h}}{\partial h^2} \right|_{h=0} \right) \right).
\end{align}
Let us consider three revealing examples. First, in a conventional superconductor, with spin-singlet s-wave pairing and spin-degenerate bands (so $\lambda_0^j\equiv0$), the $\chi^j_N(T=0) = \chi_P$ while $\chi^j_S (T=0)=0$. Second, if the same bands have a triplet pairing function characterized by the vector  $\boldsymbol{d} \perp \boldsymbol{h}$, then $\chi^j_S (0)=\chi_P + \mathcal{O} \left( \Delta^2\right)$\cite{Frigeri2004,Sigrist2009}. Third, for either of the two previous pairing scenarios, but with spin-split bands\footnote{Note that the spin-direction of Ising SOC is perpendicular to the in-plane magnetic field, otherwise the following expressions should be modified.}, $\lambda^j(\bm{k})\neq0$, the $\chi^j_S (0)=\chi_P + \mathcal{O} \left( \Delta^2\right)$\cite{Sigrist2009,Frigeri2004}. As summarized in Table~\ref{tab:scalingSOC}, the leading order of $\delta \chi^j$ (Eq.~\eqref{eq:chiOrd}) is highest (quadratic in $\Delta^j$) for triplets, or for fully split bands, giving the highest $H_{c2}$. This leading order can be lowered if the splitting vanishes at some points on the pocket (i.e., there are SOC nodes $g^j(\theta_{node})=0$ for some $\theta_{node}$, see Fig.~\ref{fig:FSspinChi}b), or even further by completely removing the splitting $\lambda_0^j=0$.
Therefore, for a fixed pocket $j$, we find a rule-of-thumb: {\it The upper critical field is lowest for the least exceptional scenario (no band splitting, singlet pairing)}.
\begin{figure}
    \includegraphics[width=\linewidth]{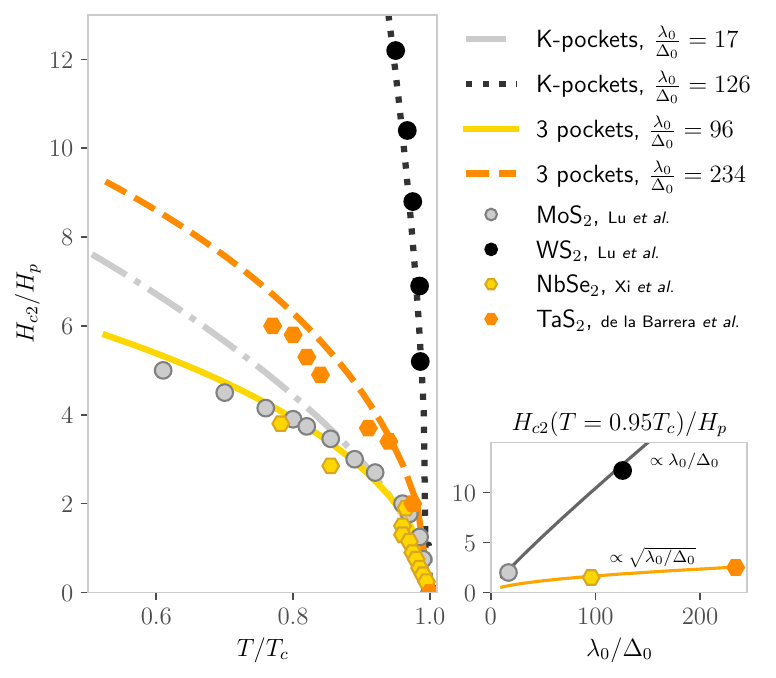}
    \caption{\label{fig:KGcomp}Calculated upper critical field compared to experimental data for monolayer Ising superconductors with a $\Gamma$-pocket (NbSe$_2$\cite{Xi2016} \& TaS$_2$\cite{DeLaBarrera2018}) and without (gated MoS$_2$\cite{Lu2015} \& WS$_2$\cite{Lu2018}). The $\Gamma$-pocket introduces a $\sqrt{\lambda_0 /\Delta_0}$-scaling of $H_{c2}$ compared to the otherwise linear $\lambda_0 /\Delta_0$-scaling in the other compounds. The ratio $\lambda_0 /\Delta_0$ is taken at the $K$-pockets.}
\end{figure}

\textit{Multipocket singlet superconductivity in TMDs.} Our key observation now is that a $\Gamma$ pocket in Ising SOC TMDs must have SOC nodes, due to the SOC nodal lines. In contrast, the $K$ pockets do not (see Fig.~\ref{fig:FSspinChi}a). Hence, for singlet pairings $\Delta^j$, reference systems with isolated pockets would have very different behaviors of $H^j_c$ (see \cite{SupMat} for definitions), raising the question of the behavior of the (unique) $H_{c2}$ of the system.

We hence solve the self-consistent gap equations (see \cite{SupMat,Aase2023}) for singlet s-wave pairings, with an additional simplifying assumption based on experimental data (at least in NbSe$_2$ \cite{Kuzmanovi2022, Khestanova2018, Valla2004}): the pairings $\Delta^\Gamma$ and $\Delta^K$ are equal in size for all temperatures near $T_c$. (We also consider adding a triplet component, which turns out to be small near $T_c$, see \cite{SupMat}.) We find that these assumptions self-consistently imply that inter-pocket pairing interaction is as strong as the intra-pocket one (see \cite{SupMat} for all details), implying that the superconductivity is strongly coupling the pockets. With such coupling, we get that very approximately:
\begin{equation}
\frac{1}{H_{c2}^2}\approx\frac{1}{3}\left(\frac{1}{(H^\Gamma_c)^2}+2\frac{1}{(H^K_c)^2}\right).
\end{equation}
If reference systems with isolated pockets $j$ would have critical fields $H_{c2}^j$ differing by a factor of at least 2, the self-consistent $H_{c2}$ of the three-pocket system would be set by the lower of the two $H_{c2}^j$. We hence extend the previous rule-of-thumb: {\it The self-consistent $H_{c2}$ of a singlet Ising SOC TMD superconductor is set by the pocket $j$ with the lowest $H_{c2}^j$, i.e., by the least exceptional part of the FS}.

\textit{Monolayer.} 
With the above insights, let us present in detail the model of monolayer NbSe$_2$ with three pockets (two $K$-pockets and one $\Gamma$-pocket of equal size). 

The $K$-pockets are modeled with a large constant $\lambda^K_0$, and $g^K(\theta)=1$.\footnote{This calculation includes the case of the nodal topological superconductor previously predicted for NbSe$_2$ under an in-plane magnetic field\protect\cite{He2018}. Within the susceptibility calculation whether the BdG bands acquire nodes when $h>\Delta^\Gamma$ or they remain gapped does not affect the form of the bands $E_{h,\zeta}$, which is the same for both cases.} One can expand in small $\Delta$ and if the integration range $\frac{\epsilon}{\Delta} \rightarrow \infty$, the difference in susceptibility becomes (see \cite{SupMat} for details)
\begin{equation} 
   \delta \chi^K (T) \approx \frac{\chi_P}{2} \frac{\Delta^2}{ \lambda_0^2} \ln  \left[ \frac{\lambda_0^2 }{\Delta^2 }\right].
\end{equation}
Thus, as known previously\cite{Ilic2017, Mockli2019, Mockli2020,Lu2015, Saito2016, Lu2018}, if only $K$-pockets are considered the critical field scales linearly with the SOC $\frac{H^K_c(T)}{H_p} \propto \frac{\lambda^K_0}{\Delta^K_0}$. In fact, for the $K$-pockets both a singlet and a triplet pairing $\Delta^K$ give $\delta \chi^{K}(T) =  \delta \chi_{2}^K(T)$.\cite{Frigeri2004}

For the $\Gamma$-pocket we model $\lambda^\Gamma (\bm{k}) = \lambda^\Gamma_0 \cos \left(3 \theta \right)$\cite{He2018,Nakata2018}, where $\theta$ is the polar angle around the $\Gamma$ point. For segment of the pocket near a SOC node (always assuming temperature near $T_c$) we determine that the difference in susceptibility is well-described by a Lorentzian $L(\theta,\beta)$ of width $\beta$ centered on the node $\theta=\theta_0$ (see \cite{SupMat} for details, and Fig.~\ref{fig:FSspinChi}b):
\begin{equation}\label{eq:dchiLor}
    \delta \chi^\Gamma(\theta, T) \approx  \left( \chi_P - \chi_{\text{sg}}(T) \right)  \frac{\pi \Delta^\Gamma_0}{\sqrt{3}\lambda^\Gamma_0} L \left(\theta, \frac{\Delta^\Gamma_0}{\sqrt{3}\lambda^\Gamma_0} \right),
\end{equation}
with $\chi_{\text{sg}}(T) =  \chi_P Y(T)$ is given by the Yoshida function $Y(T)$ characteristic of a singlet\cite{Sigrist2009}. This contribution from near a node, $\theta\approx\theta_0$, is dominant on the $\Gamma$ pocket, and we find that it closely corresponds to the numerical result for the entire pocket when $T> 0.8 T_c$ (see \cite{SupMat} for a detailed comparison). Using this expression to integrate over the entire pocket, we get
\begin{equation}
    \delta \chi^\Gamma (T)=  \left( \chi_P   - \chi_{\text{sg}}(T) \right)  \frac{\pi \Delta_0}{\sqrt{3}\lambda_0} + \mathcal{O}(\Delta_0^2),
\end{equation}
resulting in a critical field which now scales as a square root of the SOC:
\begin{equation}\label{eq:Hcsqrt}
     H^\Gamma_{c2} (T) \approx \sqrt{\frac{\lambda^\Gamma_0}{\Delta^\Gamma_0}} \sqrt{\frac{ \Omega^\Gamma_{0} (T)}{\left( \chi_P - \chi_{\text{sg}}(T) \right)  \frac{\pi }{\sqrt{3}}  }} \propto \sqrt{\frac{\lambda^\Gamma_0}{\Delta^\Gamma_0}}.
\end{equation}
Hence, for the $\Gamma$-pocket the assumption of a singlet is crucial: The SOC nodes on the $\Gamma$ pocket result in $\delta \chi^{\Gamma}(T) \gg  \delta \chi^{K}(T)$ for singlet pairing and an in-plane field.

For the full 3-pocket model of NbSe$_2$, $ \Omega_0(T)$ and $\delta \chi(T)$ are determined by the ratios $\lambda_0^K / \Delta_0^K$, $\lambda_0^\Gamma / \Delta_0^\Gamma$, and $\Delta_0^K / \Delta_0^\Gamma$. As the superconductivity in NbSe$_2$ is relatively uniform for few layers\cite{Kuzmanovi2022, Khestanova2018, Valla2004}, we have assumed that $\Delta_0^K \approx \Delta_0^\Gamma$. Due to the large SOC in these compounds, $\delta \chi^{\Gamma}(T) \gg  \delta \chi^{K}(T)$ and therefore $\delta \chi(T) \approx \delta \chi^{\Gamma}(T)$, in accord to the general discussion above. In Fig.~\ref{fig:KGcomp} measurements and calculations for NbSe$_2$ and TaS$_2$, with $ \frac{\lambda_0^\Gamma}{\Delta_0^\Gamma} \approx \frac{\lambda_0}{3 \Delta_0}$\cite{Yokoya2001,Sanders2016}, are compared to TMDs where only $K$-pockets have been considered\footnote{See \cite{SupMat} for a discussion of the additional $Q$-pockets present in gated MoS$_2$\protect\cite{Costanzo2016,Marini2023}). This compound should also be considered a multipocket compound. However, both types of pockets are split by a constant SOC and the susceptibility difference is dominated by the pockets with the lowest value.}.  

The result to highlight here is that the $\sqrt{\lambda_0 / \Delta_0}$-scaling only appears if the superconducting order parameter has a finite singlet gap at the SOC nodes. For a triplet order a $\lambda_0 / \Delta_0$-scaling is regained (see \cite{SupMat} for the critical field obtained for different pairing symmetries on a $\Gamma$-pocket). The fact that the experimental data for monolayer NbSe$_2$ and TaS$_2$ are related by $\sqrt{\lambda_0 / \Delta_0}$-scaling (Fig.~\ref{fig:KGcomp}) is therefore a strong indicator that the superconducting order is mainly a singlet at the SOC nodes.

\begin{figure}
    \centering
    \includegraphics[width=\linewidth]{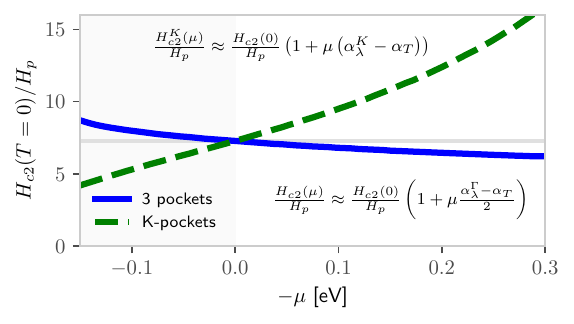}
    \caption{\label{fig:dos_doping}Prediction for the upper critical field in 1H-NbSe$_2$\cite{Rahn2012,SupMat} with a varying chemical potential $\mu$, corresponding to hole and electron doping in Eq.~\eqref{eq:allPocketsN}. $H_p$ is scaled with $T_c (\mu)$.}  
\end{figure}
\textit{Experimental prediction.} An experimental consequence of the scaling with SOC can be predicted for a doped monolayer. We note that when doping bilayer 2H-NbSe$_2$ via gating, Ref.~\cite{Xi2016a} observed a large 30\% change of $T_c$, hence gating should be a powerful probe. In our tight-binding modeling of monolayer NbSe$_2$\cite{Rahn2012}, several parameters are affected by doping. Importantly, due to the symmetry conditions on the SOC, the effective SOC at the $\Gamma$-pocket decreases under electron doping while it increases around $K$: $\lambda_0^j (\mu) \approx \lambda_0^j (0) \left( 1 + \mu \alpha_\lambda^j \right)$, with $\alpha_\lambda^\Gamma >0$ and $\alpha_\lambda^K < 0$ for electron doping $\mu <0$. Secondly, using the McMillan formula to relate $T_c$ and $\Delta_0$ as in the experimental analysis of Ref.~\cite{Xi2016a} (see \cite{SupMat} for all details), we approximate the change of pairing at low doping, $\Delta_0 (\mu) \approx \Delta_0 (0)\left( 1 + \mu \alpha_T \right)$, with $|\alpha_T|< |\alpha_\lambda^\Gamma|, |\alpha_\lambda^K|$. Finally, we extract the evolution of the density of states for each pocket $ N^j (\mu)$, as well as the $ \lambda_0^j (\mu)$, from a detailed tight-binding model, but our main prediction remains for parabolic bands with the same shifts due to $\mu$ (see \cite{SupMat}). Namely, for a singlet pairing on a FS with three pockets we find a doping-dependent critical field:
\begin{align}\label{eq:allPocketsN}
    \frac{H_{c2}(\mu)}{H_p}= \sqrt{\frac{ N^\Gamma (\mu) + 2 N^K (\mu) } { N^\Gamma (\mu)  \frac{\delta \chi^\Gamma (T,\mu)}{\chi_P - \chi_{\text{sg}}(T)} + 2 N^K (\mu) \frac{\delta \chi^K (T,\mu)}{\chi_P - \chi_{\text{sg}}(T)}}},
\end{align}
that crucially has the slope $\partial \frac{H_{c2}}{H_p}/\partial \mu  \propto (\alpha_\lambda^j - \alpha_T ) $ having opposite sign for when the $\Gamma$-pocket is included or not, as seen in Fig.~\ref{fig:dos_doping}.
Hence, also for a realistic bandstructure, we predict a linearly decreasing $H_{c2}/H_p$ with small electron doping\footnote{Electron doping of NbSe$_2$ is possible with intercalation, field effect transistors, and misfit structures. Hole doping, while possible, is more difficult to achieve for large doping values, due to the high work function in TMDs\protect\cite{Zullo2023}.}, or correspondingly increasing with hole doping, as a clear signature of the $\sqrt{\lambda_0^\Gamma / \Delta_0}$-scaling due to the dominant role of the $\Gamma$-pocket, while a linearly increasing trend is predicted in absence of the $\Gamma$-pocket.

\textit{Discussion and conclusions.} We have demonstrated the necessity of including all bands when calculating the upper critical field of a superconductor. The critical field can be limited by the sections of the FS where $\bm{g}$ or $\bm{d}$  is no longer perpendicular to the field. We have illustrated this principle by comparing $H_{c2}$ for Ising SOC superconductors either having or not pockets where the SOC has nodal points. We have shown a huge difference in $H_{c2}$ arising for different pairing symmetries at these few SOC-node points. Therefore, the behavior of the upper critical field could be used to indicate which pairing symmetries at specific points in momentum space are favored.

For monolayer 1H-NbSe$_2$ and 1H-TaS$_2$, our analysis shows that a dominant triplet order parameter is inconsistent with the experimental data. This rules out any purely $f$-wave\cite{Shaffer2020} or nematic order\cite{Cho2022,Roy2025,Siegl2024}. Furthermore, even though our calculation only includes a uniform singlet order, other proposed orders for NbSe$_2$ are consistent with our results. This includes a nodal singlet, a topologically non-trivial singlet order such as a $d +id$-wave, or the recently proposed nodal topological superconductor\cite{He2018, Shaffer2020,Cohen2024, Margalit2021}. A mixed parity order, $s + f$-wave\cite{Das2023, Wan2022} or a $d_y$-triplet component appearing under a field\cite{Tang2021,Mockli2020,Kuzmanovi2022}, would also only affect the subleading order of $\delta \chi (T)$ and thus only slightly change the value of $H_{c2}$. It should also be reiterated that the signature from the $K$-pockets are similarly almost entirely hidden, as their potential influence on $H_{c2}$ is of the same size as disorder effects\cite{Ilic2017, Mockli2020}. 

Other works have also found the $\Gamma$-pocket essential to accurately describe a pairing mechanism dependent on spin-fluctuations\cite{Siegl2024}, reinforcing the conclusion that the nodal points can play a vital role. 

Further, our conclusions are consistent with the available data on few-layer TMDs with 2H-stacking\footnote{This will be discussed in detail elsewhere.}. In homobilayers, the overall inversion symmetry is preserved. However, the SOC within each layer has nodal points which should be taken into account when calculating $H_{c2}$.

Note that a Rashba SOC $\lambda_R$, e.g., due to a substrate, lifts the Ising SOC nodes. We show that the correction to the susceptibility difference is only of order $(\lambda_R/\lambda_0)^2$, and hence the influence on $H_{c2}$ is negligible (see \cite{SupMat}).

In this work, we focused on monolayer Ising SOC superconductors as characteristic inversion-symmetry-breaking materials with only a few points in the FS without a SOC splitting. However, the idea is general as it relies on symmetries and shows that the critical field could similarly be used to extract information on the nature of the order parameter in hetero- and/or Moiré structures.

Finally, our approach might be used to investigate the nature of pairing in Ising superconductors whose bandstructure and/or superconductivity has been altered by a wide range of experimental control parameters, such as pressure or strain (\cite{Abdel-Hafiez2016, Henriquez-Guerra2024}), electron- and hole-doping via field-effect gating or the introduction of an external electric field\cite{Liu2017, Ye2012}. Notably, the family of misfit dichalcogenide compounds, formed by the stacking of rocksalt and TMD layers, offers an efficient way to tune the effective TMD chemical potential, potentially to large values.

\textit{Acknowledgment.} We thank fruitful discussions with M. Calandra, C. Quay and I. Paul.
This work was supported by the French Agence Nationale de la Recherche (ANR), under grant number
ANR-22-CE30-0037.

\bibliography{HC_ising_G_refs}

\end{document}


\title{Suppl. Material to ``Upper critical field and pairing symmetry of Ising superconductors''}

\author{Lena Engstr\"om}
\affiliation {Université Paris-Saclay, CNRS, Laboratoire de Physique des Solides, 91405 Orsay, France}
\author{Ludovica Zullo}
\affiliation{
Institut für Theoretische Physik und Astrophysik and Würzburg-Dresden Cluster of Excellence ct.qmat,
Universität Würzburg, 97074 Würzburg, Germany}
\author{Tristan Cren}
\affiliation{Sorbonne Université, CNRS, Institut des Nanosciences de Paris, UMR7588, F-75252 Paris, France}
\author{Andrej Mesaros}
\affiliation{Université Paris-Saclay, CNRS, Laboratoire de Physique des Solides, 91405 Orsay, France}
\author{Pascal Simon}
\affiliation{Université Paris-Saclay, CNRS, Laboratoire de Physique des Solides, 91405 Orsay, France}

\maketitle

\section{Effective monolayer model}\label{sec:effModel}
To derive an effective model for a $\Gamma$-pocket we calculate the susceptibility for a single order parameter and expand the expressions close to a node of the spin-orbit coupling (SOC). We consider an approximately circular pocket around $\bm{k}=0$, with a parabolic dispersion $\xi_{\bm{k}} = \mu - \alpha k^2 $. The bands are spin polarized and split by the spin-orbit coupling $\lambda(\theta) = \lambda_0 \cos 3 \theta$, such that the dispersion of each band in the normal state is $\xi_\zeta = \xi_{\bm{k}} + \zeta \lambda(\theta)$, where $\zeta=\pm 1$. With spin-singlet pairing and an in-plane magnetic Zeeman field $h$ (along the $x$-axis), the BdG Hamiltonian is
\begin{equation}
    \mathcal{H}_{\bm{k}} = \xi_{\bm{k}} \sigma_0 \otimes \tau_z + \lambda(\theta) \sigma_z \otimes \tau_z + h \sigma_x \otimes \tau_0 + \Delta \sigma_0 \otimes \tau_x
\end{equation}
with $\sigma_i$ and $\tau_i$, $i=0,x,y,z$, being the Pauli matrices in spin and particle-hole space, respectively, using the basis $\{ c_{\bm{k} \uparrow}, c_{\bm{k} \downarrow},$ $ c_{-\bm{k} \downarrow}^\dagger, -c_{-\bm{k} \uparrow}^\dagger \}$. The BdG bands are:
\begin{equation}
    E_{\zeta,h} = \sqrt{\xi_k^2 + h^2 +  |\lambda (\theta)|^2 + \Delta^2 + 2 \zeta \sqrt{h^2 (\xi_k^2 + \Delta^2) + \xi_k^2  |\lambda (\theta)|^2 }}
\end{equation}
and $ \xi_{\zeta,h}= \xi_{ k} +\zeta \sqrt{ |\lambda (\theta)|^2 + h^2}$. The derivatives in field of the normal state and superconducting bands are simply
\begin{equation}
  \left. \frac{\partial \xi_{\zeta,h}}{\partial h} \right|_{h=0} = \left. \frac{\partial E_{\zeta,h}}{\partial h} \right|_{h=0} =0
\end{equation}
\begin{equation}
  \left. \frac{\partial^2 \xi_{\zeta,h}}{\partial h^2} \right|_{h=0} = \frac{\zeta}{|\lambda (\theta)|}, \qquad \left. \frac{\partial^2 E_{\zeta,h}}{\partial h^2} \right|_{h=0} =  \frac{\zeta}{ |\lambda (\theta)|} \left(  \frac{\xi_\zeta}{E_\zeta} +  \frac{\Delta^2}{\xi_k E_\zeta}\right)
\end{equation} 
At $h=0$ the bands are given by $E_{\zeta} = \sqrt{ \xi_\zeta ^2 + \Delta^2 }$. For a parabolic dispersion the density of states are constant and are set to that of the FS $N(0)$. Integration in the radial direction $k$ results in a normal state susceptibility\cite{Frigeri2004, Sigrist2009}:
\begin{align}\label{eq:chiN}
\chi_{N}(T) = \sum_{\zeta, \bk}  \left. \frac{\partial^2 \xi_{\zeta,h}}{\partial h^2} \right|_{h=0} f( \xi_{\zeta})  =  2 \sum_{\zeta}  \frac{N(0)}{2 \pi}  \int d \theta \int_{k_\text{min}}^{k_\text{max}} dk \frac{k \zeta }{|\lambda (\theta)|} f( \xi_{\zeta})\\ \notag
 = \frac{N(0)}{2 \pi}  \int d \theta \frac{ 1 }{|\lambda (\theta)|} \left(  T \ln \left(\frac{ f ( \epsilon_\text{max} + \lambda (\theta)) f ( \epsilon_\text{min} - \lambda (\theta))}{f ( \epsilon_\text{max} - \lambda (\theta)) f ( \epsilon_\text{min} + \lambda (\theta))} \right)   \right) = 2 N(0)= \chi_P
\end{align}
where $f(\xi)$ is the Fermi-Dirac function at temperature $T$. The limits of the integration range is set to $k_\text{min}^2 = \frac{\mu - \epsilon_\text{min}}{\alpha} $ and $k_\text{max}^2 = \frac{\mu - \epsilon_\text{max}}{\alpha} $. As discussed in the main text, we assume $\epsilon_\text{min}=- \epsilon_\text{max}=-\epsilon$ and $\epsilon > \lambda_0$. The interband susceptibility in the superconducting state can be divided into four terms, via $\chi_{s} = \chi_{s,T} + \chi_{s,0}$ where
\begin{equation}\label{eq:chiS}
    \chi_{s,T} =  \sum_{\zeta, \bm{k}} \left. \frac{\partial^2 E_{\zeta,h}}{\partial h^2} \right|_{h=0}  f(E_{\zeta}), \qquad \chi_{s,0} =  \frac{1}{2} \sum_{\zeta, \bm{k}} \left( \left. \frac{\partial^2 \xi_{\zeta,h}}{\partial h^2} \right|_{h=0} - \left. \frac{\partial^2 E_{\zeta,h}}{\partial h^2} \right|_{h=0} \right).
\end{equation}
\begin{figure}
    \centering
    \includegraphics[width=0.5\linewidth]{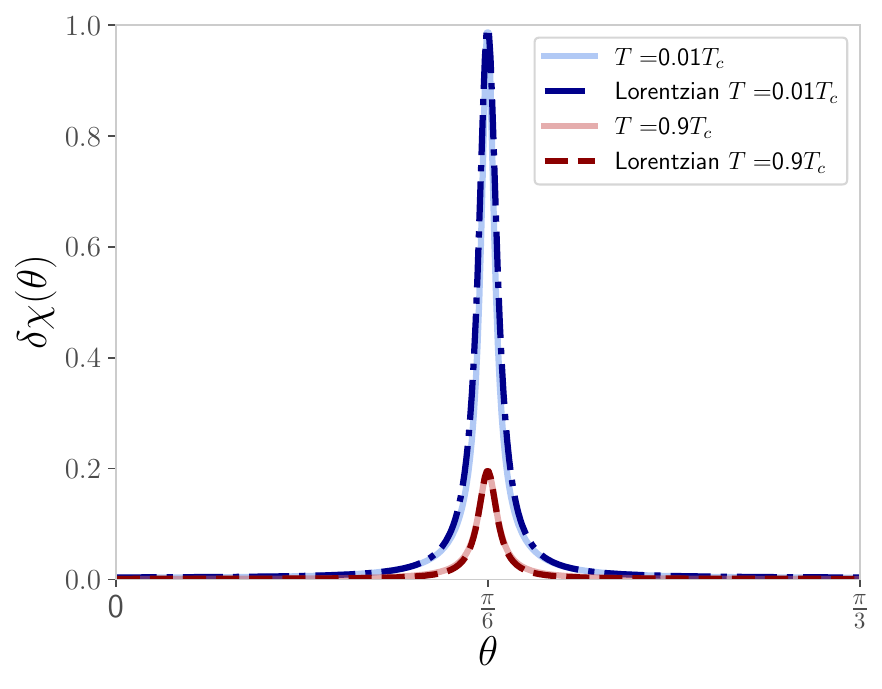}
    \caption{\label{fig:peaksComp}The susceptibility difference $\delta \chi (\theta) = \chi_N -  \chi_S (\theta) $ in a region around a node in the SOC on the $\Gamma$-pocket, forms a peak for a singlet order. The effective model, approximating the peak with a Lorentzian, Eq.~\eqref{eq:LorGen}, is shown along with the numerical integration of the susceptibility for $\lambda_0 / \Delta_0= 32$. The peak becomes sharper for an increased $\lambda_0 / \Delta_0$, while furthest from the node $\delta \chi (\theta=0) \propto (\Delta_0 / \lambda_0 )^2$, as the bands are fully split at that point.}
\end{figure}
A further separation is made into $\chi_{s,0} = \chi_{s,0,1} + \chi_{s,0,2}$ and $\chi_{s,T} = \chi_{s,T,1} + \chi_{s,T,2}$. In section~\ref{sec:TripNodes} we show that the contributions of $\chi_{s,0,1}$, $\chi_{s,T,1}$ are recovered when considering a $d_z$-triplet pairing. For a singlet pairing and a parabolic dispersion of the bands, the terms are calculated as
\begin{align}
\chi_{s,0,1} = \sum_{\zeta} \frac{N(0)}{2 \pi}  \int d \theta \int d \xi_k
\frac{\zeta}{ |\lambda (\theta)|} \left( 1-  \frac{\xi_\zeta}{E_\zeta} \right) \\ \notag
= -\frac{N(0)}{2 \pi}  \int d \theta \frac{ 1}{|\lambda (\theta)|} \left( - \sqrt{(\epsilon + |\lambda (\theta)| )^2 + |\Delta|^2} + \sqrt{(\epsilon - |\lambda (\theta)| )^2 + |\Delta|^2} \right)
\end{align}
\begin{align}
   \chi_{s,0,2}= \sum_\zeta  \frac{N(0)}{2 \pi} \int d \theta \int d \xi_k \frac{\zeta \Delta^2}{|\lambda (\theta)| \xi_k E_\zeta}\\ \notag
    = -\frac{N(0)}{2 \pi} \int d \theta \frac{\Delta^2}{|\lambda (\theta)|  \sqrt{\Delta^2 + |\lambda (\theta)| ^2}} \left[ \text{arctanh} \left( \frac{|\lambda (\theta)|^2 + |\lambda (\theta)|  \epsilon + \Delta^2 }{\sqrt{\Delta^2 + |\lambda (\theta)| ^2} \sqrt{\left( \epsilon + |\lambda (\theta)|\right)^2 + \Delta^2 }}\right)  \right.\\ \notag
    \left. -  \text{arctanh} \left( \frac{|\lambda (\theta)|^2 - |\lambda (\theta)|  \epsilon + \Delta^2 }{\sqrt{\Delta^2 + |\lambda (\theta)| ^2} \sqrt{\left( \epsilon - |\lambda (\theta)|\right)^2 + \Delta^2 }}\right) \right]
\end{align}
\begin{equation}
\chi_{s,T,1} =  2 \sum_{\zeta} \frac{N(0)}{2 \pi} \int d \theta \int d \xi_k \frac{ \zeta }{|\lambda (\theta)|}  \frac{\xi_{\zeta}}{\sqrt{\xi_{\zeta}^2 + |\Delta|^2}} f( E_{\zeta})
= \frac{N(0)}{2 \pi} \int d \theta \frac{ 1 }{|\lambda (\theta)|} \left(  T \ln \left(\frac{ f ( E_{+,\text{max}}) f ( E_{-,\text{min}})}{f ( E_{-,\text{max}}) f ( E_{+,\text{min}})} \right)   \right) + \chi_{s,0}
\end{equation}
\begin{equation}
   \chi_{s,T,2}= \sum_\zeta \frac{N(0)}{2 \pi} \int d \theta \int d \xi_k  \frac{\zeta \Delta^2}{|\lambda (\theta)| \xi_k E_\zeta}  f(E_\zeta)
\end{equation}
To determine the effect of a node in the SOC, at some $\tilde{\theta}=0$, we expand the terms around $\lambda (\theta) \approx v_\lambda \tilde{\theta}$, with $v_\lambda = 3 \lambda_0$:
\begin{equation}~\label{eq:chi0sing}
   \chi_{s,0} (\theta) \approx \frac{N(0)}{2 \pi} \left(  \frac{2 \epsilon}{ \sqrt{\Delta^2 + \epsilon^2}} - \frac{2 \epsilon}{ \sqrt{\Delta^2 + \epsilon^2}}   + (v_\lambda \tilde{\theta})^2 \frac{2 \epsilon ( 3 \Delta^2 + \epsilon^2)}{ 3 \Delta^2 (\epsilon^2 + \Delta^2)^\frac{3}{2}} \right) = \alpha_\theta \tilde{\theta}^2
\end{equation}
Clearly $\chi_{s,0}(\tilde{\theta}=0) =0$. For $\chi_{s,T}$ we expand around the node before performing the integration in the radial direction $k$. The lowest order terms are:
\begin{equation}
   \chi_{s,T,1} \approx  \frac{N(0)}{2 \pi} \int d \xi_k \frac{2 }{(\xi_k^2 + \Delta^2 )^{3/2}} \left(   \Delta^2 f (\sqrt{\xi_k^2 + \Delta^2}) - \xi_k^2 \sqrt{\xi_k^2 + \Delta^2} \frac{e^{\sqrt{\xi_k^2 + \Delta^2}/T}}{\left( 1 + e^{\sqrt{\xi_k^2 + \Delta^2}/T} \right)^2 T} \right) 
\end{equation}
\begin{equation}
   \chi_{s,T,2} \approx  \frac{N(0)}{2 \pi} \int d \xi_k  \frac{-2 \Delta^2}{(\xi_k^2 + \Delta^2 )^{3/2}} \left(  f (\sqrt{\xi_k^2 + \Delta^2}) + \frac{e^{\sqrt{\xi_k^2 + \Delta^2}/T}}{\left( 1 + e^{\sqrt{\xi_k^2 + \Delta^2}/T} \right)^2 T}\right) 
\end{equation}
Combining the terms, we can determine that for any temperature at the node
\begin{equation}\label{eq:chiNode}
   \chi_{s}  (\tilde{\theta}=0)= - N(0)\int d \xi_k  \frac{\partial f(E_0)}{\partial E_0} =  Y(T) \chi_P
\end{equation}
where $E_0 = \sqrt{\xi_k^2 + \Delta^2}$. $Y(T) $ is the Yoshida function, where $Y(0) \approx \sqrt{\frac{2 \pi \Delta_0}{T}} e^{-\Delta_0 /T}$ and $Y(T_c) =1$. We identify Eq.~\eqref{eq:chiNode} as being just the susceptibility of a band-degenerate singlet superconductor.  

\begin{figure}
    \includegraphics[width=0.32\linewidth]{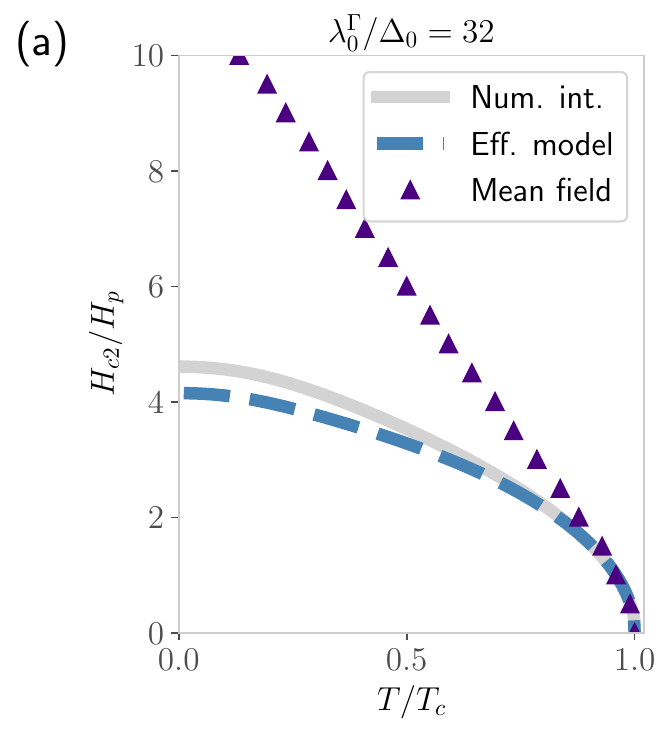}
    \includegraphics[width=0.32\linewidth]{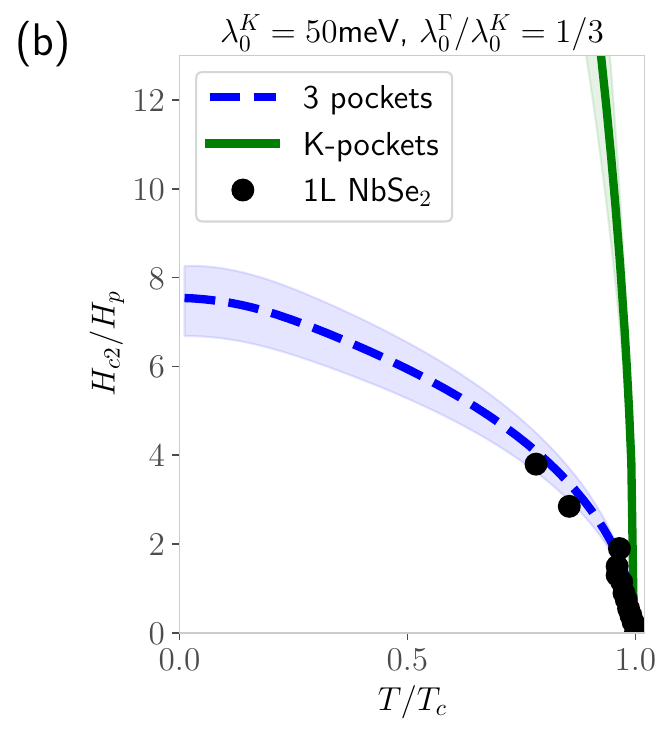}
    \includegraphics[width=0.32\linewidth]{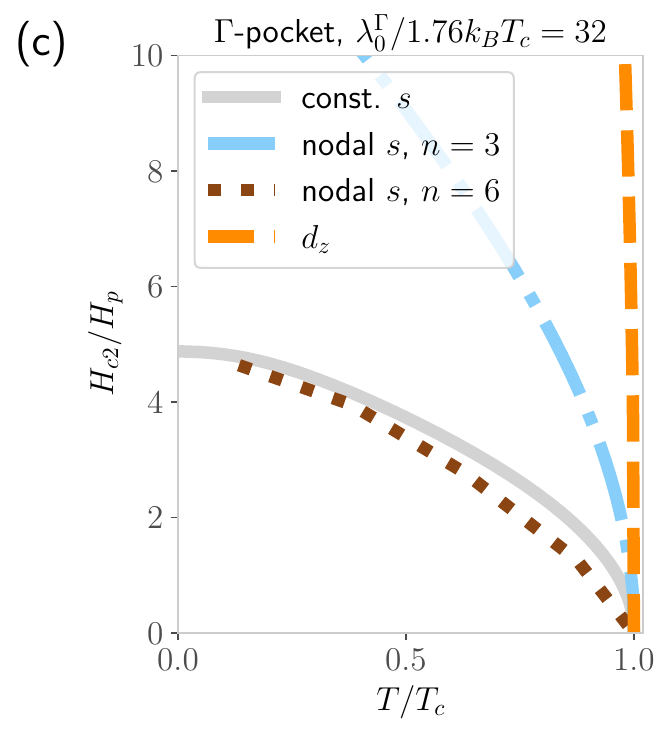}
    \caption{\label{fig:comp3K}a) For a single $\Gamma$-pocket the effective Lorentzian model is compared to the critical field given by the numerical evaluation of the susceptibility. The critical field determined by a self-consistent mean-field solution, for a uniform singlet pairing on a $\Gamma$-pocket, is compared to the two other solutions. b) For monolayer NbSe$_2$ the singlet pairing $H_{c2}$ is compared for the model with all 3 pockets and for when only the $K$-pockets are included. Around each curve the results for a 20\% variation of $\lambda_0^K$ is shown. c) For a $\Gamma$-pocket the critical field is compared for different pairing symmetries. A nodal singlet is given by $\Delta(\theta) \approx \Delta \cos (n \theta )$, where the choice $n=3$ is an example showing the node corresponding to a constant ratio $\Delta (\theta) / \lambda (\theta)$. The triplet is constant in the region and calculated numerically via Eq.~\eqref{eq:chidz}.}
\end{figure}
\subsection{Lorentzian approximation near a SOC node}
As argued in the main text, the susceptibility difference is large at the SOC nodes and of order $(\Delta_0 / \lambda_0 )^2$ far from the node. Fig.~\ref{fig:peaksComp} shows the angular dependence of the derived expressions $\delta \chi (\theta) = \chi_N -  \chi_S (\theta)$ around a node on a $\Gamma$-pocket, where $\chi_S = \int d \theta \chi_S (\theta) / 2 \pi$. The susceptibility difference can be modeled as a Lorentzian peak centered on the node at $\theta_0$:
\begin{equation}\label{eq:LorGen}
    L(\theta,\beta) = \frac{1}{\pi} \frac{\beta}{(\theta-\theta_0)^2 + \beta^2}
\end{equation}
The Lorentzian parameters are approximated from $\chi_{s,0}(\theta) \approx \alpha_\theta (\theta - \theta_0)^2$, calculated in the previous section. With the expansion $L(\theta,\beta) \approx \frac{1}{\beta \pi} -  \frac{(\theta - \theta_0)^2}{\beta^3 \pi}$ and Eq.~\eqref{eq:chi0sing} we identify $\beta \approx \frac{\Delta_0}{\sqrt{3}\lambda_0}$ and scale the function by $ \frac{\pi \Delta_0}{\sqrt{3}\lambda_0}$. In Fig.~\ref{fig:comp3K}a) the critical field is compared for $\Delta \chi$ calculated via numerical integration and the effective Lorentzian model. For a high SOC and $T>0.8 T_c$ the effective model is valid while at lower temperatures the discrepancy increases.

\subsection{Derivation of the susceptibility around the K-pockets}
For a pocket with a constant SOC $\lambda (\theta) = \lambda_0$, the smallest quantity in the superconducting susceptibility is the superconducting pairing $\Delta$. Expanding in this parameter gives us
\begin{align}
\chi_{s,0,1} \approx - \frac{N(0)}{2 \pi} \int d \theta \frac{ 1}{|\lambda (\theta)|} \left( 2(\epsilon -  |\lambda (\theta)| )  - 2(\epsilon +  |\lambda (\theta)| )  + \Delta^2 \left( - \frac{1}{\epsilon +  |\lambda (\theta)| } + \frac{1}{\epsilon -  |\lambda (\theta)| }  \right) \right) \\ \notag
=  \frac{N(0)}{2 \pi} \int d \theta \left( 2  -  2 \frac{\Delta^2 }{\epsilon^2 -  |\lambda (\theta)|^2 }  \right)
\end{align}
and the leading term of the susceptibility becomes
\begin{equation} 
   \chi_{s,0}  \approx  \chi_P - 2 N(0) \left( \frac{\Delta^2}{\epsilon^2 - \lambda_0^2} - \frac{\Delta^2}{ \lambda_0^2} \frac{1}{2}\left( - \ln  \left[ \frac{2\lambda_0^2 (\epsilon + \lambda_0)^2}{ \Delta^2 \epsilon^2} - 1\right] - \ln  \left[ \frac{2\lambda_0^2 (\epsilon - \lambda_0)^2}{\Delta^2 \epsilon^2} - 1\right] \right) \right).
\end{equation}
If we allow $\epsilon / \Delta \rightarrow \infty$ all that remains is
\begin{equation} 
   \chi_{s}  \approx  \chi_P  - \frac{2 N(0)}{2} \frac{\Delta^2}{ \lambda_0^2} \ln  \left[ \frac{\lambda_0^2 }{\Delta^2 } \right] 
\end{equation}
In a superconductor which does not break inversion symmetry the integration limit $\epsilon$ is only required to be larger than the superconducting order parameter and is often set to the Debye frequency. 
As established in the main text, a SOC requires that $\epsilon > \lambda_0$, as to include all uncompensated states in the bands further from the FS. 

In Fig.~\ref{fig:comp3K}b) the 3-pocket model for NbSe$_2$ is compared to if only the $K$-pockets were included in the calculation. By varying the value of $\lambda_0^K$, and consequentially $\lambda_0^\Gamma$, to include different reported values from ARPES\cite{Yokoya2001,Nakata2018,Sanders2016} and DFT\cite{Wickramaratne2023, DeLaBarrera2018} we observe a stark difference between the two models.

\subsection{Triplet or nodal singlet pairing}\label{sec:TripNodes}
As has been previously established, for a triplet pairing, parallel to the direction of the SOC, the order remains purely intraband for any in-plane field $h$, with $E_{\zeta,h} = \sqrt{\xi_{\zeta,h}^2 + \Delta^2 }$ \cite{Sigrist2009}. The susceptibility for a $d_z$-triplet is thus determined by
\begin{equation}\label{eq:chidz}
    \chi_S^{d_z} = \chi_{s,T,1} + \chi_{s,0,1} 
\end{equation}
At a $K$-pocket, keeping the integration range $\epsilon$ finite, we obtain the leading term
\begin{equation}
   \chi_{s,0}^{d_z}  \approx  \chi_P  -2 N(0)\frac{\Delta^2}{\epsilon^2 - \lambda_0^2} 
\end{equation}
For simplicity, the triplet pairing has here been kept to a constant gap $\Delta$. For a more accurate evaluation the order parameter should have a symmetry in momentum space with odd parity, e.g. an $f$-wave pairing.

If the pairing at the $\Gamma$-pocket is a triplet the effect of the SOC nodes is greatly reduced. The susceptibility for a $d_z$-pairing is given by Eq.~\eqref{eq:chidz}. Around a node the susceptibility can be evaluated via numerical integration for such a triplet, results shown in Fig.~\ref{fig:comp3K}c). The $d_z$-triplet is kept gaped at the node. As there no longer are points on the FS which are have a more conventional order than the rest, the susceptibility at the nodes have the same quadratic scaling with the order parameter $\delta \chi_{2}$, as all other points of the FS. 

To fully determine the importance of the pairing at the nodal points we have also calculated the susceptibility for a singlet with nodes coinciding with the SOC nodes. Close to a node both the SOC $\lambda(\theta) \approx v_\lambda (\theta - \theta_0)$ and the pairing $\Delta(\theta) \approx v_\Delta (\theta - \theta_0)$ can be expanded in the angle. A singlet must have an even parity order parameter $\Delta(\theta) \approx \Delta \cos (n \theta )$, where $n$ is an even integer. Expanding $\chi (\theta)$ for $n=6$ close to the node results in a lowest order term:
\begin{equation}
    \chi_S  (\theta) \approx N(0) \int d \theta \int d \xi_{\bm{k}} 2 \frac{\partial f (\xi_{\bm{k}})}{\partial \xi_{\bm{k}}} +  \mathcal{O} \left( \theta^2 \right) = \chi_P +  \mathcal{O} \left( \theta^2 \right)
\end{equation}
As there is no superconducting order parameter at the node, $\Delta(\theta_0)=0$, the normal state susceptibility is regained at this point. For $n=3$, the ratio $\Delta (\theta ) / \lambda (\theta )$ remains constant at all $\theta$, and coinciding nodes have a minimal effect on the susceptibility. In Fig.~\ref{fig:comp3K}c) the critical field is less suppressed than for the gaped singlet. We expect the susceptibility difference to be of order $\delta \chi_{2}$, and numerically $H_{c2}$ scales linearly with SOC. However, the most realistic nodal singlet order in NbSe$_2$ should have even parity and a 3-fold symmetry, thus given by $n=6$. In Fig.~\ref{fig:comp3K}c) we can observe that for this order the critical field is similarly suppressed as for the uniform singlet. In this case the states close to the node have a very small SOC and a finite superconducting order, as the ratio $\Delta (\theta ) / \lambda (\theta )$ varies greatly along the FS.  The size of $\Delta (T)$ is determined by $T_c$, which is set to be the same for each pairing symmetry, and thus the ratio $\lambda_0^\Gamma / \Delta_0$ is not fixed. Numerically we establish that for $n=6$ the $H_{c2}$ also scales as $\sqrt{\lambda_0^\Gamma / \Delta_0}$. Hence, we conclude that only a realistic singlet order close to the SOC nodes results in a square root dependence of SOC. The possibility of a nodal versus uniform singlet is explored in the following section~\ref{sec:MF}, using mean-field theory. 

\section{Self-consistent mean-field solution}\label{sec:MF}
\begin{figure}
    \includegraphics[width=0.4\linewidth]{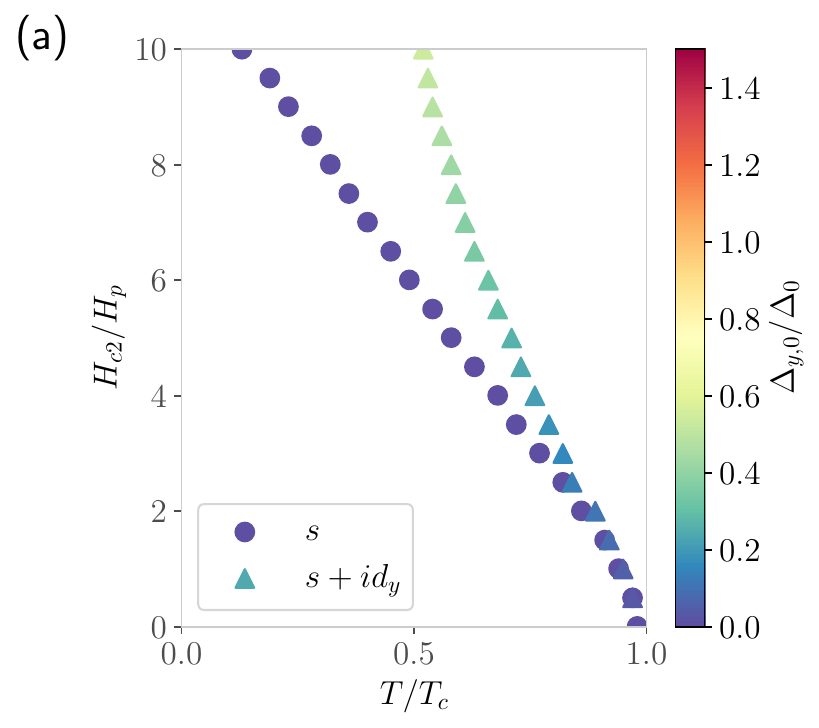}
    \includegraphics[width=0.4\linewidth]{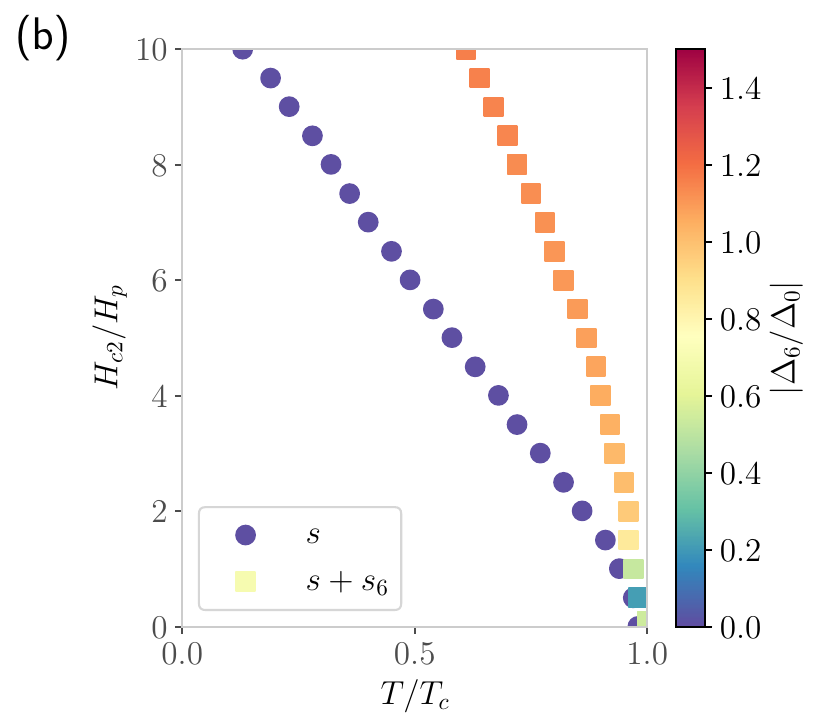}
    \caption{\label{fig:compMF}Mean-field solutions for pairing symmetries beyond a uniform singlet, for $\lambda_0 / \Delta_0= 32$. a) $s + i d_y$ pairing, Eq.s~\eqref{eq:MFs} \& \eqref{eq:MFdy}. The coupling parameter for the triplet order is set to a large value of $\sum_{k} g_y^2 (\bm{k})V_y /N_k V = 1$. Even for this favoring of a triplet order, the order parameter remains small at $T>0.8T_c$. b) The singlet order parameter is allowed to be a mix between a uniform and a nodal $\Delta_{\text{nodal}} = \Delta_6 \cos (6 \theta )$  order, where once again the coupling strength for the second order is $\sum_{k} g_6^2 (\bm{k})V_6 /N_k V = 1$. At low fields $\Delta_6 <0$.}
\end{figure}
The susceptibility accurately describes the evolution of the thermodynamic potential under a magnetic field given that the field is small. We therefore want to determine limits to the validity of calculating the critical field via the susceptibility. By solving a mean-field self-consistency equation for the order parameter $\Delta(h,T)$, the critical field can be defined as the $H_{c2} (T)$ at which the order parameter goes to zero. To compare the validity of the models we have calculated $H_{c2}$ on a single $\Gamma$-pocket with a parabolic dispersion and $\lambda (\theta) = \lambda_0 \cos (3 \theta)$. 

As demonstrated by Ref.~\cite{Mockli2019}, a field-induced $d_y$-triplet component of the pairing is favorable if NbSe$_2$ is a mainly singlet superconductor. To include this effect the self-consistency calculations were performed for two cases: for only a singlet order parameter and for the possibility of a mixed-parity order. If there is a uniform $s$-wave singlet $\Delta$ and an odd parity $i d_y$ triplet $\Delta_y$ (both real-valued) there are two bands
\begin{equation}
    E_{\pm}^{\text{s}+ i \text{d}_y} = \sqrt{\xi_k^2 + h^2 + \lambda_\bk^2 + |\Delta|^2 + |\Delta_y|^2 \pm 2 \eta_\bk}
\end{equation}
with $\eta_\bk =\sqrt{h^2 (\xi_k^2 + |\Delta|^2) +2 \Delta \Delta_y h \lambda_\bk+ (  \xi_k^2 + |\Delta_y|^2) \lambda_\bk^2 } $. The triplet $\Delta_y = \Delta_{y,0} g_y (\bm{k})$, is an odd function of momentum where we choose to study the $f$-wave pairing $g_y (\bm{k}) = \cos (3 \theta)$. In a standard way, one self-consistency equation is found for each order parameter $\Delta_\eta$ at an extrema in the free energy functional $\frac{\partial \Omega (h,T)}{\partial \Delta_{\eta}} =0$. By determining that
\begin{equation}
    \frac{\partial E_{\pm}^{\text{s}+ i \text{d}_y}}{\partial \Delta} = \frac{\Delta \pm \frac{h (h \Delta + \lambda_\bk\Delta_y)}{\eta_\bk}}{E_{\pm}^{\text{s}+ i \text{d}_y}}, \qquad  \frac{\partial E_{\pm}^{\text{s}+ i \text{d}_y}}{\partial \Delta_{y,0}} = \frac{\Delta_{y,0} g_y^2 (\bm{k}) \pm \frac{g_y (\bm{k}) \lambda_\bk(h \Delta + \lambda_\bk \Delta_y)}{\eta_\bk}}{E_{\pm}^{\text{s}+ i \text{d}_y}}
\end{equation}
the two coupled self-consistency equations become 
\begin{equation}\label{eq:MFs}
    \Delta = - \frac{V_s}{ N_k} \sum_\bk \left[  \frac{\Delta + \frac{h (h \Delta + \lambda_\bk \Delta_y)}{\eta_\bk}}{2 E_{+}^{\text{s}+ i \text{d}_y}}  \tanh \left( \frac{E_{+}^{\text{s}+ i \text{d}_y}}{2 T}\right) +  \frac{\Delta - \frac{h (h \Delta + \lambda_\bk\Delta_y)}{\eta_\bk}}{2 E_{-}^{\text{s}+ i \text{d}_y}}  \tanh \left( \frac{E_{-}^{\text{s}+ i \text{d}_y}}{2 T}\right) \right]
\end{equation}
\begin{align}\label{eq:MFdy}
    \Delta_{y,0} = - \left( \frac{N_k}{\sum_{\bk'} g_y^2 (\bm{k}')} \right) \frac{V_y}{ N_k} \sum_\bk \left[  \frac{\Delta_{y,0} g_y^2 (\bm{k}) + \frac{g_y (\bm{k}) \lambda_\bk(h \Delta + \lambda_\bk\Delta_y)}{\eta_\bk}}{2 E_{+}^{\text{s}+ i \text{d}_y}}  \tanh \left( \frac{E_{+}^{\text{s}+ i \text{d}_y}}{2 T}\right) \right.\\ \notag
    \left. +  \frac{\Delta_{y,0} g_y^2 (\bm{k}) - \frac{g_y (\bm{k}) \lambda_\bk(h \Delta + \lambda_\bk\Delta_y)}{\eta_\bk}}{2 E_{-}^{\text{s}+ i \text{d}_y}}  \tanh \left( \frac{E_{-}^{\text{s}+ i \text{d}_y}}{2 T}\right) \right]
\end{align}
for $N_k$ points in the BZ. Effective interactions are required to be attractive, $V,V_y <0$, for non-trivial solutions to be possible. For each order parameter $\Delta_{\sigma, \sigma'} (\bm{k})= \Delta_{\sigma, \sigma'}^{(0)} g_{\sigma, \sigma'} (\bm{k})$ we define the effective interaction on the form $V_{\sigma, \sigma'} (\bm{k}, \bm{k}') = V_{\sigma, \sigma'} g_{\sigma, \sigma'} (\bm{k})  g_{\sigma, \sigma'} (\bm{k}') $, for spin labels $\sigma, \sigma' = \uparrow, \downarrow$. This allows us to express the order parameter as
\begin{equation}
   \Delta_{\sigma, \sigma'}^{(0)} g_{\sigma, \sigma'} (\bm{k}) = - \frac{1}{N_k}  g_{\sigma, \sigma'} (\bk) \sum_{\bm{k}'} V_{\sigma, \sigma'} g_{\sigma, \sigma'} (\bm{k}')  \langle c_{-\bm{k}', \sigma} c_{\bm{k}', \sigma'} \rangle
\end{equation}

In Fig.~\ref{fig:comp3K}a) the $H_{c2} (T)$ determined by the self-consistency equations where only a singlet was allowed is compared to the susceptibility calculations. The effective interaction $V$ was chosen to give $T_c=3$K at $h=0$. For temperatures $T>0.8 T_c$ the methods have overlapping results. However, the mean-field solution has a much larger $H_{c2}$ than that of the susceptibility calculation at low temperatures. Known differences between the two calculations are that the mean-field calculation includes higher order terms in field $h$ and that the order parameter is a function of the field $\Delta (h,T)$\cite{Tang2021,Kuzmanovi2022}, which is not included in the susceptibility calculation. From the comparison of the two methods, it is reasonable to assume that the susceptibility calculation can only be considered valid close to $T_c$.

Including the possibility of a triplet pairing allows us to determine how well the assumption of a uniform spin-singlet is as the field increases. Many studies have attempted to determine the superconducting order in monolayer NbSe$_2$. In previous works the self-consistent equations for the gap have been solved, involving all spin-symmetries as well as disorder\cite{Mockli2020,Tang2021,Siegl2024,Roy2025}. To determine how large the effects from other pairing symmetries could be we have considered two combinations of order parameters in Fig.~\ref{fig:compMF}, with a uniform singlet with either the $d_y$-triplet or a nodal singlet. The nodal singlet order 
$\Delta_{\text{nodal}} = \Delta_6 \cos (6 \theta )$ has an effective coupling $V_6$. Even when the coupling parameter $V_i$ for the additional pairing is large, a superconductivity with nodes coinciding with that of the SOC is not favored. In the range of temperatures in which the susceptibility calculation is valid, we can thus conclude that a gaped singlet is a likely order. Of note for the mean-field calculation performed in this work is the ignored effect of the $K$-pockets for the pairing. However, previous works including more realistic 3-pocket models similarly find the singlet order to be dominant\cite{Mockli2018,Wickramaratne2023,Das2023}.

\section{Self-consistent $H_{c2}(T)$ for multi-pocket superconductor}\label{sec:multiPocket}
In our susceptibility calculations with multiple pockets, we have implicitly made certain assumptions about the coupling between the pockets. The superconducting order parameter has been treated as a shared parameter for the pockets. However, removing the $\Gamma$-pocket or removing the $K$-pockets would result in systems with very different critical fields. Hence, it is important to consider how the critical magnetic field of a multi-pocket system depends on the strength of the inter-pocket coupling. In this section we explore the physical motivation behind our implicit assumptions via a self-consistent calculation for three pockets and a uniform singlet order parameter.

To deal with the three pockets, we introduce two coupled order parameters, $\Delta_\Gamma (T,h)$ and $\Delta_K (T,h)$. In accordance with experimental data\cite{Khestanova2018,Kuzmanovi2022} that shows the same gap size at these pockets, we impose that the order parameters are equal at $h=0$: $\Delta_\Gamma (T,0)= \Delta_K (T,0)$. The additional physical consideration taken into account is that close to $T_c$ we expect one common $H_{c2}$ for both pockets\cite{Cho2022}, again consistent with experiments. We initially keep the inter-pocket coupling $g_{\Gamma K}$ general in the two coupled self-consistent gap equations:
\begin{align}
    \Delta_\Gamma (T,h) = g_\Gamma \Delta_\Gamma (T,h) F_\Gamma (T,h, \Delta_\Gamma (T,h)) + 2 g_{\Gamma K} \Delta_K (T,h) F_K (T,h, \Delta_K (T,h))\\ \notag
        \Delta_K (T,h) = 2 g_K \Delta_K (T,h) F_K (T,h, \Delta_K (T,h)) + g_{\Gamma K} \Delta_\Gamma (T,h) F_\Gamma (T,h, \Delta_\Gamma (T,h)),
\end{align}
where $g_K,g_\Gamma$ are intra-pocket coupling parameters and the factor 2 comes from the doubled density of states due to having two $K$-pockets. Since the order parameters are uniform in momentum, they have been moved outside of the $F_j$-functions, which are:
\begin{align}
    F_j(T,0,\Delta_j) = \frac{1}{2} \int d \epsilon d\theta \left( \frac{1}{2 E_{j,+}} \tanh{\frac{E_{j,+}}{2 T} } +  \frac{1}{2 E_{j,-}} \tanh{\frac{E_{j,-}}{2 T} } \right)
\end{align}
\begin{align}
    F_j(T,h,\Delta_j) = \frac{1}{2}  \int d \epsilon d\theta \left( \frac{1 + \frac{h^2}{\eta_{j,\bk}}}{2 E_{j,+,h}} \tanh{\frac{E_{j,+,h}}{2 T} } +  \frac{1- \frac{h^2}{\eta_{j,\bk}}}{2 E_{j,-,h}} \tanh{\frac{E_{j,-,h}}{2 T} } \right).
\end{align}
The bands $E_{j=K,\sigma,h}$ and $\eta_{j=K, \bm{k}}$ use a SOC $\lambda_{\bm{k}} = \lambda_0^K$ for $j=K$, while those for $j=\Gamma$ use $\lambda_{\bm{k}} = \lambda_0^\Gamma \cos 3 \theta$, and these bands are introduced in section \ref{sec:MF}. The functions $F_j$ have been assumed to only have contributions from states close of the FS and no inter-pocket order parameter is included (note, $\Gamma-K$ inter-pocket pairing would have a non-zero center-of-mass momentum, like FFLO pairing). The coupling parameters are related due to the shared temperature dependence of $\Delta_j$ having one shared $T_c$:
\begin{align}\label{eq:selfTc}
    1 = g_\Gamma  F_\Gamma (T_c,0,0) + 2 g_{\Gamma K} F_K (T_c,0,0)\\ \notag
        1 =2 g_K  F_K (T_c,0,0) + g_{\Gamma K} F_\Gamma (T_c,0,0)
\end{align}
An additional simplifying approximation, which we find to be well supported by numerical integration, is that for a uniform singlet order parameter $ F_\Gamma (T,0, 0) \approx  F_K (T,0, 0)$, making the condition for the critical temperature in Eq.~\ref{eq:selfTc} become:
\begin{align}\label{eq:selfTh0}
    1  = g_{\text{tot}} F_K (T_c,0,0),
\end{align}
where $ g_{\text{tot}} \equiv  g_\Gamma + 2 g_{\Gamma K} = 2 g_K + g_{\Gamma K} $. If we now fix the temperature to some $T$ close to $T_c$ and consider a non-zero $h$, which can now be assumed to be small, we can Taylor expand:
\begin{align}\label{eq:scExpand}
    \Delta_\Gamma (T,0) = g_\Gamma \Delta_\Gamma (T,0) \left(  F_\Gamma (T,0, \Delta_\Gamma (T,0)) +  F_\Gamma^{''} h^2 \right) + 2 g_{\Gamma K} \Delta_K (T,0) \left(  F_K (T,0, \Delta_K (T,0)) +  F_K^{''} h^2 \right)\\ \notag
    \Delta_K (T,0) = 2g_K \Delta_K (T,0) \left(  F_K(T,0, \Delta_K (T,0)) +  F_K^{''} h^2 \right) + g_{\Gamma K} \Delta_\Gamma (T,0) \left(  F_\Gamma (T,0, \Delta_\Gamma (T,0)) +  F_\Gamma^{''} h^2\right),
\end{align}
where we have defined $ F_j^{''}= \left. \frac{\partial^2 F_j (T, h, \Delta_j)}{\partial h^2} \right|_{h=0}$.

We come to a key point of this analysis. As we aim to understand the shared $H_{c2}$ in terms of the behavior of the critical fields associated with isolated pockets, we now need to define precisely the reference system that has an isolated pocket (more precisely, either only a $\Gamma$-pocket, or only the $K$ and $-K$ pocket pair). First, note that without the pocket coupling ($g_{\Gamma K}=0$), each pocket would have its separate critical field $H_{c2}^\Gamma (T), H_{c2}^K (T)$. We note that like in the susceptibility calculation, the critical fields that are comparable are those that share a $T_c$. That means that we must define a reference system having an isolated pocket in such a way that its critical temperature is equal to the unique $T_c$ of the three-pocket system. Another way to understand this issue is by starting from the three-pocket system and imagining a removal, e.g., of the $K$-pockets. This would remove their contribution from the susceptibility but also from the condensation energy, hence changing the critical temperature. So, the $H_{c2}^\Gamma$ we need to define is not necessarily the critical field one would get if, for example, the system was doped so that the other $K$-pocket no longer was present at the FS. We instead want to more meaningfully define a reference $\Gamma$-pocket-only system by removing the contribution of the $K$-pockets from the susceptibility, but preserving their condensation energy so as to preserve the value of $T_c$. Hence the critical fields of the separate reference systems $j=\Gamma,K$ will have to be defined by $H_{c2}^j (T=T_c) =0$. To achieve this, the separate self-consistency equations for the isolated pockets are defined by using rescaled coupling parameters $\tilde{g}_K, \tilde{g}_\Gamma$:
\begin{align}\label{eq:hcDecoup}
    1= \tilde{g}_\Gamma  \left(  F_\Gamma (T,0, 0) +  F_\Gamma^{''} (H_{c2}^\Gamma)^2 \right) \\ \notag
    1 = \tilde{g}_K \left(  F_K(T,0, 0) +  F_K^{''} (H_{c2}^K)^2 \right)
\end{align}
where from here on $ F_j^{''}=  F_j^{''}(T,0,0)$. As the density of states of each pocket is assumed equal for simplicity and $ F_\Gamma (T,0, 0) \approx  F_K (T,0, 0)$, the coupling parameters are equal $\tilde{g} \equiv \tilde{g}_\Gamma = \tilde{g}_K$. The critical temperature is given by the condition $1= \tilde{g} F_K (T_c,0, 0)$, which by our setup must correspond to the $T_c$ in Eq.~\eqref{eq:selfTh0}. We thus require that $\tilde{g} = g_{\text{tot}} $. 

\begin{figure}
    \centering
    \includegraphics[width=0.4\linewidth]{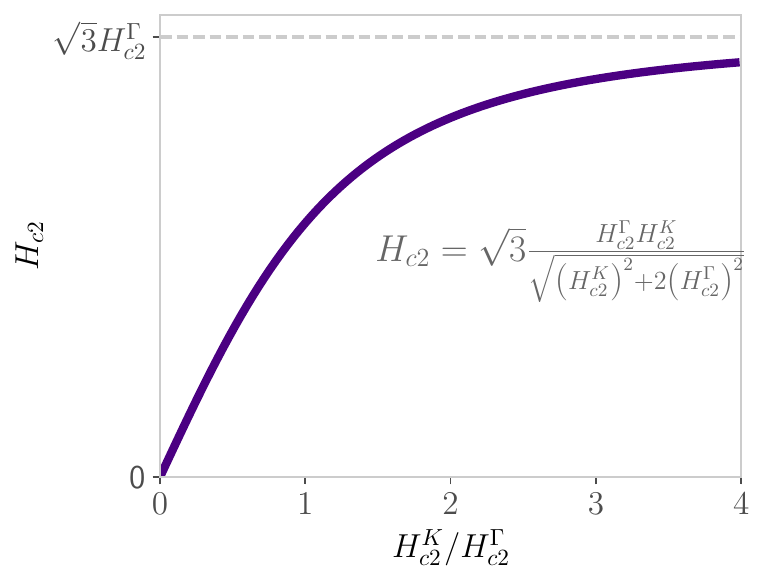}
    \caption{\label{fig:PowMean}The critical field $H_{c2}$ as given by three coupled pockets $g_{\Gamma K} = g_K =g_\Gamma$, as in Eq.~\eqref{eq:1sol}, where if they were decoupled one would have $H_{c2}^\Gamma$ and the other two $H_{c2}^K$.}
\end{figure}

To solve for the $H_{c2}$ of the coupled pockets, we plug the found relations into Eq.~\eqref{eq:scExpand} and linearize the self-consistency equations, by setting $ F_j (T,0, \Delta_j ) =  F_j (T,0, 0 )$. The two solutions to the critical field are then
\begin{align}\label{eq:2sol}
      H_{c,1}= \frac{H_{c2}^\Gamma H_{c2}^K}{\sqrt{\left(  H_{c2}^K\right)^2 +  \frac{2 g_{\Gamma K}}{g_\Gamma} \left( H_{c2}^\Gamma\right)^2}} \sqrt{ \frac{g_\Gamma + 2 g_{\Gamma K}}{g_\Gamma}} ,  \qquad  H_{c,2}= \frac{H_{c2}^\Gamma H_{c2}^K}{\sqrt{ \frac{2 g_{ K}}{g_{\Gamma K}}\left(  H_{c2}^\Gamma\right)^2 +  \left( H_{c2}^K \right)^2}} \sqrt{ \frac{2g_K +  g_{\Gamma K}}{g_{\Gamma K}}} 
\end{align}
It is immediately clear that if we were to set $g_{\Gamma K}=0$, where the pockets are uncoupled, we recover the two different critical fields $H_{c,1}= H_{c2}^\Gamma$ and $H_{c,2}= H_{c2}^K$. However, to find the unique self-consistent critical field of the three-pocket system we must require $H_{c2}\equiv H_{c,1}\equiv H_{c,2}$. This is possible only if $g_{\Gamma K} = g_K = g_\Gamma$. Note that our assumption $\Delta_K (T,0) = \Delta_\Gamma (T,0)$ has imposed restrictions on the coupling between the pockets. The critical field for the coupled pockets is thus:
\begin{align}\label{eq:1sol}
     H_{c}(T)= \sqrt{3}\frac{H_{c2}^\Gamma (T) H_{c2}^K (T)}{\sqrt{\left(  H_{c2}^K (T)\right)^2 + 2 \left( H_{c2}^\Gamma (T) \right)^2}} =M_{-2} (H_{c2}^\Gamma (T), H_{c2}^K (T), H_{c2}^K (T)),
\end{align}
where the standard power mean of the $n=3$ variables is:
\begin{align}
    M_p (\{x_j \}) = \left(\frac{1}{n} \sum_{j=1}^n x_j^p \right)^{1/p}.
\end{align}
In Fig.~\ref{fig:PowMean} the resulting critical field is plotted as a function of the ratio $\frac{H_{c2}^K (T)}{H_{c2}^\Gamma (T)}$, where $H_{c2}$ is linear up until $\frac{H_{c2}^K (T)} {H_{c2}^\Gamma (T)}\approx 1$ and plateaus after $\frac{H_{c2}^K (T)} {H_{c2}^\Gamma (T)}\approx 4$.

When one of the critical fields is much lower than the other $\frac{H_{c2}^\Gamma (T)}{H_{c2}^K (T)} \ll 1$:
\begin{align}
    H_{c} (T) \approx \sqrt{3} H_{c2}^\Gamma (T ) \left( 1 -  \left( \frac{H_{c2}^\Gamma (T )}{H_{c2}^K (T )} \right)^2 \right).
\end{align}
To lowest order the $H_{c2}$ hence becomes simply the $H_{c2}^\Gamma$, up to the factor $\sqrt{3}$ which stems directly from our definition of the reference $\Gamma$-pocket-only system as having only one pocket (instead of three) but still the same $T_c$.

The final result can now be compared to our susceptibility calculation, where the equivalent setup has one $\Gamma$-pocket and two $K$-pockets, with $ \Omega_0^\Gamma (T) =  \Omega_0^K (T) $ while $\delta \chi^\Gamma (T) \gg \delta \chi^K (T) $:
\begin{align}
    H_{c2} (T) = \sqrt{\frac{\Omega_0^\Gamma (T) +2 \Omega_0^K (T)}{\delta \chi^\Gamma (T) + 2\delta \chi^K (T)}} \approx  \sqrt{\frac{3 \Omega_0^\Gamma (T)}{\delta \chi^\Gamma (T) }}  \left(1 - \frac{\delta \chi^K(T)}{\delta \chi^\Gamma (T)} \right)  = \sqrt{3} H_{c2}^\Gamma (T ) \left( 1 - \left( \frac{H_{c2}^\Gamma (T )}{H_{c2}^K (T )} \right)^2 \right),
\end{align}
proving that the two methods are equivalent close to $T_c$, and that our susceptibility calculation implicitly assumed a strong inter-pocket coupling $g_{\Gamma K} \approx g_{K}\approx g_\Gamma$.

As the restrictions set in this problem requires the order parameters to have a common $T_c$ and $H_{c2}$, it effectively reduces to a one-gap problem. In general, the coupling parameters will depend on the pairing mechanism, making a solution to the multi-gap self-consistency equations not necessarily the minima of the free energy\cite{Aase2023}. If we introduce a small anisotropy between the pockets $T_{c,K} \gtrapprox T_{c,\Gamma}$, we expect that the attractive coupling parameters are no longer exactly equal $|g_\Gamma| \gtrapprox |g_K|$, which results in both $|g_{ K} | \gtrapprox |g_{\Gamma K}|$ and $|g_{ \Gamma} |\gtrapprox |g_{\Gamma K}|$. Under these assumptions the solution where both parameters have the same phase $\theta_K = \theta_\Gamma$, where $\Delta_j (T,H) = |\Delta_j (T,H)| e^{i \theta_j}$, is the global minimum to the two-gap-problem at $g_k \approx g_\Gamma \approx g_{\Gamma K }$\cite{Aase2023}.

\section{Prediction for doping monolayer}
\begin{figure}
    \centering
    \includegraphics[width=0.5\linewidth]{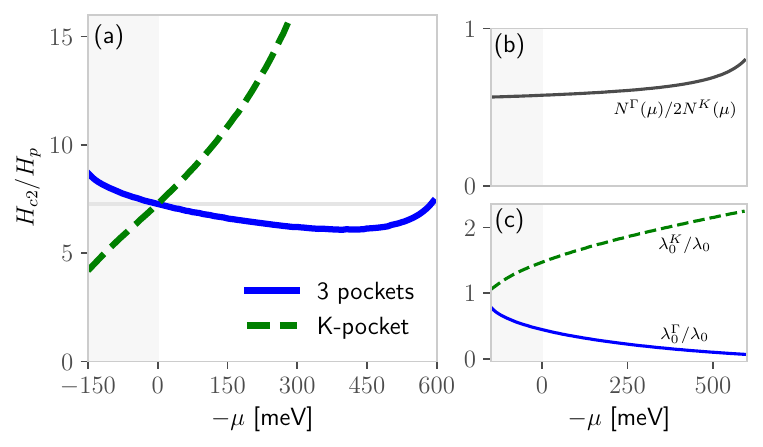}
    \includegraphics[width=0.33\linewidth]{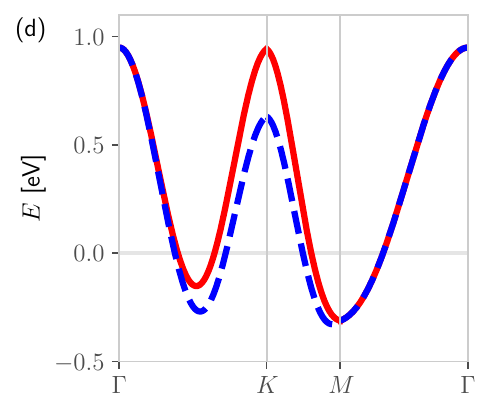}
    \caption{\label{fig:dos_doping}a) The upper critical field for hole and electron doped NbSe$_2$ up to higher values of the chemical potential $\mu$ than shown in the main text. The Pauli limit $H_p$ is at each point scaled to the predicted $T_c (\mu)$ for the density of states $N(\mu)$. b) The proportion of density of states at the FS belonging to $\Gamma$-pocket remains around $N^\Gamma (\mu) /2 N^K (\mu) =0.6$ for $|\mu| < 0.2$eV. c) For electron doping the effective SOC at the $\Gamma$-pocket decreases while it increases at the $K$-pockets. d) The bandstructure for Eq.~\eqref{eq:real_bands}, fit to ARPES\cite{Rahn2012}, with $\lambda_0 =60$meV.}  
\end{figure}
We predict that the different scaling behaviors of the models, either including or excluding the $\Gamma$-pocket, can be distinguished under doping. In a realistic model of the bands in the monolayer 1H-MX$_2$ compounds, like given by Ref.~\cite{Rahn2012} for NbSe$_2$, the one-band tight-binding model has several parameters that are affected by doping. In bilayer 2H-NbSe$_2$ Ref.~\cite{Xi2016a} could dope the compound via gating, resulting in a 30\% change of $T_c$. At multiple doping values, the $T_c (\mu)$ and total density of states at the FS $N(\mu) = N^\Gamma (\mu) + 2 N^K (\mu)$ was measured and found to follow a strong coupling formula relating the two quantities. To allow us to predict the evolution of the in-plane critical field $H_{c2}$, there are three important quantities that change within the accessible doping range:
\begin{enumerate}
    \item Critical temperature $T_c (\mu)$ and thus $\Delta_0^K  (\mu) =\Delta_0^\Gamma (\mu) =\Delta_0 (\mu) = 1.76 k_{\text{B}} T_c (\mu)$. 
    \item Magnitude of SOC at each pocket $\lambda_0^j (\mu)$, $j=\Gamma,K$.
    \item Proportion of the total DoS at the FS coming from the $\Gamma$-pocket $ N^\Gamma (\mu) /2 N^K (\mu)$.
\end{enumerate}
As we will see, the third point becomes negligible at small doping. Of note is that even though the $T_c (\mu)$ could be approximated at a given doping from a model of the bandstructure it can also be measured in experiment.

The tight-binding model from Ref.~\cite{Rahn2012} is shown in Fig.~\ref{fig:dos_doping} and is described by:
\begin{align}\label{eq:real_bands}
    E(\bm{k}) =& t_0+\mu_0 + t_1 (2 \cos \xi \cos \eta + \cos 2 \xi) \\ \notag
    & + t_2 (2 \cos 3 \xi \cos \eta + \cos 2 \eta) \\ \notag
    & + t_3 (2 \cos 2 \xi \cos 2 \eta + \cos 4 \xi) \\ \notag
    & + t_4 (\cos \xi \cos 3 \eta + \cos 5 \xi \cos \eta + \cos 4 \xi \cos 2 \eta ) \\ \notag
    & + t_5 (2 \cos 3 \xi \cos 3 \eta + \cos 6 \xi) 
\end{align}
with $\xi = \frac{1}{2} k_x a$, $\eta = \frac{1}{2} \sqrt{3} k_y a$ and the values $(t_0, t_1, t_2, t_3, t_4, t_5 )$ $=(203.0, 46.0, 257.5, 4.4, -15.0, 6.0 )$meV. The SOC is assumed to be the lowest harmonic respecting the full symmetry of the BZ:
\begin{align}\label{eq:SOCsym}
    \lambda (\bm{k}) = 2 \lambda_0 \sin \xi ( \cos \xi - \cos \eta)
\end{align}
where we choose $\mu=-150$meV and $\lambda_0 = 60$meV to fit the FS of undoped monolayer 1H-NbSe$_2$ for $E_\zeta (\bm{k}) = E (\bm{k}) + \zeta \lambda (\bm{k})$.

By symmetry, $|\lambda (\bm{k})|$ will be maximal around the $K$-points and necessarily goes to zero at the $\Gamma$-point. Expanded around the $\Gamma$-point $\lambda^\Gamma_k (\mu) \approx \lambda_0  k_\mu ^3 \cos 3 \theta= \lambda_0^\Gamma (\mu ) \cos 3 \theta$, and doping around $k_\mu = k_F + \delta k_\mu$ gives us a $\lambda_0^\Gamma (\mu )  = \lambda_0^\Gamma (0) \left( 1+ \mu \alpha_\lambda^\Gamma \right)$. Around the $K$-pocket 
\begin{equation}
  \lambda_k^K (\mu) \approx \left( -2 \cos \frac{2 \pi}{\sqrt{3}} (-1 + \sin \frac{2 \pi}{\sqrt{3}} ) + \cos \frac{2 \pi}{\sqrt{3}} (-1 + 4  \sin \frac{2 \pi}{\sqrt{3}}) q_\mu^2 \right) \lambda_0 
\end{equation}
and $\lambda^K_k (\mu) \approx  \lambda_0^K (0) \left( 1 + \mu \alpha_\lambda^K \right)$, where $q$ is the radial momentum from the $K$-point. Expanding Eq.~\eqref{eq:real_bands} around the high symmetry points $\Gamma$ and $K$ we find small trigonal warping corrections beyond the approximation of the bands as parabolic:
\begin{align}
    E^K(\bm{q}) \approx \alpha_0^K  +q \alpha_1^K   \cos \theta +  (\alpha_2^K  + \alpha_{2c}^K  \cos (2 \theta)) q^2\\
     E^\Gamma(\bm{k}) \approx \alpha_0^\Gamma +  \alpha_2^\Gamma k^2
\end{align}
where $\alpha_{1}^K/\alpha_{2}^K < 0.1$, and $\alpha_{2c}^K/\alpha_{2}^K < 0.01$. Keeping only the parabolic dispersion, we find $\alpha_\lambda^K = -\cos \frac{2 \pi}{\sqrt{3}} (-1 + 4  \sin \frac{2 \pi}{\sqrt{3}}) \frac{\lambda_0}{\alpha_2^K \lambda_0^K (0)} $ and $\alpha_\lambda^\Gamma = \frac{3}{2} \sqrt{ \frac{-\alpha_0^\Gamma}{(\alpha_2^\Gamma )^3} }$ with the numerical values $\alpha_\lambda^K  \approx -1$ and $\alpha_\lambda^\Gamma  \approx 3.9$.

In Ref.~\cite{Allen1975} $T_c$ is connected to the DoS via the strong coupling formula:
\begin{eqnarray}\label{eq:strCoup}
    T_c (\mu) = \frac{\omega_{\text{log}}}{1.2} e^{- \frac{1.04 (1+ \bar{\lambda} (\mu))}{\bar{\lambda} (\mu) - \mu^\ast (1 + 0.62 \bar{\lambda}(\mu))}}
\end{eqnarray}
Here $\bar{\lambda}(\mu) = V N(\mu)$, where $V$ is the electron-phonon coupling constant and $\mu^\ast$ is the Coulomb pseudopotential. From fitting the values $\mu^\ast =0.1$ and  $\bar{\lambda}(0) =1.1$ has been determined to describe bilayer 2H-NbSe$_2$\cite{Xi2016a}. For a small doping $\mu$ we consider the expansion $T_c (\mu) \approx T_c (0) \left( 1  + \mu \alpha_T \right)$ where $\alpha_T \approx -1.34$. 

We expect the effect from the Coulomb potential to have a different value in the monolayer. However, we find that the difference cannot be sufficiently large to negate the predicted experimental signature. By comparing undoped monolayer NbSe$_2$ to the misfit superconductor (LaSe)$_{1.14}$(NbSe$_2$), we can use these two data points to evaluate the change in critical temperature and doping. In the misfit structure, separated monolayers of NbSe$_2$ are superconducting with $T_c =1.23$K and are doped with $\mu \approx - 0.32$eV\cite{Samuely2023}. A roughly approximated value $\alpha_T^{\text{mono}} \approx -1.84$ might therefore be appropriate for monolayers.

To calculate $H_{c2}$ of the realistic bands, in Eq.~\eqref{eq:real_bands}, with an assumed singlet pairing: 
\begin{align}\label{eq:allPocketsN}
    \frac{H_{c2}(\mu)}{H_p}= \sqrt{\frac{ N^\Gamma (\mu) + 2 N^K (\mu) } { N^\Gamma (\mu)  \frac{\delta \chi^\Gamma (T,\mu)}{\chi_P - \chi_{\text{sg}}(T)} + 2 N^K (\mu) \frac{\delta \chi^K (T,\mu)}{\chi_P - \chi_{\text{sg}}(T)}}} 
\end{align}
 the DoS $ N^j (\mu)$ and the SOC $ \lambda_0^j (\mu)$ are extracted for the bands at a given doping $\mu$. In Fig.~\ref{fig:dos_doping} the calculated $H_{c2}$ is compared for all 3 pockets to a calculation using only the $\lambda_0^K (\mu)$ of the $K$-pockets (when $N^\Gamma (\mu)=0$). The two models have a different scaling with $\mu$ originating from the symmetry of the SOC. As can be seen in Fig.~\ref{fig:dos_doping}b the ratio $N^\Gamma (\mu) /2 N^K (\mu)$ can be considered constant at $|\mu| <0.2$eV. For the full model the scaling can be expanded as
\begin{equation}
    \frac{H_{c2} (\mu)}{H_p} \propto \sqrt{\frac{\lambda_0^\Gamma (\mu)}{\Delta_0 (\mu)}} \sqrt{\frac{N (\mu)}{N_\Gamma (\mu)}} \approx  \sqrt{\frac{\lambda_0^\Gamma (0)}{\Delta_0 (0)}} \sqrt{\frac{N (0)}{N_\Gamma (0)}}\sqrt{\frac{1+ \alpha_\lambda^\Gamma \mu}{1+ \alpha_T \mu}}
\end{equation}
while if only the $K$-pockets are considered
\begin{equation}
    \frac{H_{c2}^K (\mu)}{H_p} \propto \frac{\lambda_0^K (\mu)}{\Delta_0 (\mu)} \approx  \frac{\lambda_0^K (0)}{\Delta_0 (0) } \left(1 + \mu \left( \alpha_\lambda^K - \alpha_T \right) \right).
\end{equation}
The slope $\partial \frac{H_{c2}}{H_p}/\partial \mu$ has opposite sign for the two models, as seen in Fig.~\ref{fig:dos_doping}a, which is described by
\begin{equation}\label{eq:scalingDop}
       \frac{H_{c2} (\mu)}{H_p} \approx  \frac{H_{c2} (0)}{H_p} \left(1 + \mu \frac{\alpha_\lambda^\Gamma - \alpha_T}{2} \right), \qquad  \frac{H_{c2}^K (\mu)}{H_p} \approx  \frac{H_{c2}^K (0)}{H_p} \left(1 + \mu \left( \alpha_\lambda^K - \alpha_T \right)  \right)
\end{equation}
where $\alpha_\lambda^\Gamma >0$ and $\alpha_\lambda^K < 0$ for electron doping $\mu <0$, while $|\alpha_T|< |\alpha_\lambda^\Gamma|, |\alpha_\lambda^K|$. Our prediction therefore becomes that if monolayer 1H-NbSe$_2$ has a scaling of $H_{c2}$ determined by the $\Gamma$-pocket the critical field (normalized by the Pauli limit) decreases under electron doping. It should here once again be noted that this scaling arises when the pairing symmetry is predominately singlet.
\begin{figure}
    \centering
    \includegraphics[width=0.35\linewidth]{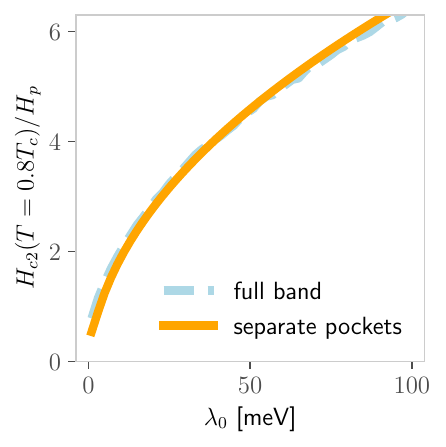}
    \caption{\label{fig:num_full_bands}Upper critical field calculated from the susceptibility via numerical integration, Eq.s~\eqref{eq:chiN} \& \eqref{eq:chiS}, for the bands in Eq.~\eqref{eq:allPocketsN}, assuming a singlet order. The bands are integrated numerically in a grid of $1500 \times 1500$ $k$-points and compared to the susceptibility calculated for separated parabolic bands with $\lambda_0^\Gamma = 0.4 \lambda_0$ and $N^\Gamma (\mu) /2 N^K (\mu) =0.6$.}
\end{figure}

In Eq.~\eqref{eq:allPocketsN} the simplified model where the susceptibility for each pocket can be calculated separately is still assumed. To determine the validity of this assumption we have numerically integrated a singlet order $E_\Delta (\bm{k}) = \sqrt{E_\zeta (\bm{k})^2 + \Delta}$ for the bands $E_\zeta (\bm{k})$ in Eq.~\eqref{eq:real_bands}. In Fig.~\ref{fig:num_full_bands} the $H_{c2}$ is shown for an increasing SOC $\lambda_0$ and compared to the effective model in Eq.~\eqref{eq:allPocketsN}. Even though the numerical integration includes terms beyond a parabolic dispersion the curves from the two methods closely overlap and follow the same scaling. Excluded from this model is separate coupling terms for the different pockets, only allowing for a uniform singlet order as in section~\ref{sec:multiPocket}.

\section{Ising and Rashba SOC}
In many experimental set-ups the presence of an additional Rashba SOC is expected, due to effects such as ripples in the monolayer\cite{Huertas-Hernando2006}. Due to symmetry constraints the SOC in a one-orbital model must be given by a function $\hat{\lambda} (\bm{k} )= \vec{g} (\bm{k}) \cdot \vec{\sigma}$, where $\vec{g} (\bm{k}) = - \vec{g} (-\bm{k}) $ is antisymmetric\cite{Samokhin2009}. However, any nodal point $|\hat{\lambda} (\bm{k} )|= 0$ must be given by the symmetry group of the problem. For Ising SOC the vector has the form $\vec{g} (\bm{k}) = (0,0,\lambda_I (\bm{k}))$, such that the function $\hat{\lambda} (\bm{k} )$ always has nodal lines connecting $\bm{k}=0$ to the edge of the BZ. Expanding the function around the origin, it can be expressed as $\lambda_I (\bm{k}) = \sum_n C_n (k) \cos (n \theta + b_n(k))$, where only antisymmetric functions in $\theta$ are allowed, fixing $n$ to be an odd integer. Each nodal line must also be connected to a second nodal line via time reversal symmetry. In the symmetry group $C_{3v}$ of our model of the monolayer TMDs, the full antisymmetric SOC is always some linear combination of Ising and Rashba SOC on the form\cite{Samokhin2009}: $\hat{\lambda} (\bm{k} ) = \lambda_R \left( k_y \sigma_x - k_x \sigma_y \right) + \lambda_I \left( k_+^2 + k_-^2\right) \sigma_z$ with $k_\pm = k_x \pm i k_y$.

An effect of the addition of a small Rashba SOC strength $\lambda_R$ is that it lifts the Ising SOC nodes, and consequentially nodal points at the FS, of the Ising SOC $\lambda_I ( \theta)$. We define the Rashba SOC on a circular pocket as
\begin{equation}
    \hat{\lambda}_R (\theta) = \lambda_R \left( \sin \theta \hat{x} - \cos \theta \hat{y}\right) 
\end{equation}
The same form of the Rashba SOC can be used for both $\Gamma$- and $K$-pockets. The Rashba SOC given in the full $D_{3h}$ symmetry of a layer of NbSe$_2$\cite{He2018} has the form $\hat{\lambda}_R (\bm{k}) = V_R \left[ \sqrt{3} \sin \left( \frac{\sqrt{3} k_y}{2} \right) \cos \left( \frac{ k_x}{2} \right) \sigma_x - \left( \sin k_x \cos \left( \frac{\sqrt{3} k_y}{2} \right)  \right) \sigma_y \right]$. Expanding this expression close to the high symmetry points gives us the approximate value $\lambda_R^K = - \lambda_R^\Gamma /2$.

From the Rashba SOC, the bands obtain an additional in-plane spin-polarization with a momentum-dependent direction. At each node of the Ising SOC, the normal state bands $\xi_\zeta$ have a fully in-plane spin-polarization $\sin \theta_0 \hat{x} - \zeta \cos \theta_0 \hat{y}$. For a $\Gamma$-pocket these points occur at $\theta_0 = (2n -1)\frac{\pi}{6}$, with $n=1, \dots, 6$.

The normal state bands $\xi_\zeta$ and the bands with either a singlet or $d_z$-triplet order $E_\zeta$ take the form:
\begin{equation}
    \xi_\zeta = \xi_k + \zeta \lambda(\theta), \qquad   E_\zeta = \sqrt{\xi_\zeta^2 + \Delta^2} 
\end{equation}
where $ \lambda(\theta)= \sqrt{ \lambda_R^2 + |\lambda_I (\theta)|^2 } $. The in-plane magnetic field can be chosen along one direction $\hat{h} = h \hat{x} $, as the in-plane critical field ends up independent of the field angle under these approximations. The normal state becomes:
\begin{align}
    \xi_{\zeta, h} = \xi_k + \zeta \sqrt{  \left( \lambda_R  \sin (\theta) + h \right)^2 + \lambda_R^2  \cos^2 (\theta) + |\lambda_I (\theta)|^2 }
\end{align}
The SOC can be divided into the components which are parallel and perpendicular to the field:
\begin{align}
    \lambda_\parallel (\theta) = \lambda_R |\sin (\theta)|, \qquad 
 \lambda_\perp (\theta) = \sqrt{ \lambda_R^2 \cos ^2 (\theta) + |\lambda_I (\theta)|^2 }\\
    \lambda(\theta)= \sqrt{\lambda_\perp^2 (\theta) + \lambda_\parallel ^2(\theta) } = \sqrt{ \lambda_R^2 + |\lambda_I (\theta)|^2 }
\end{align}
where we from here on out use $\lambda_R >0$. Only a Rashba contribution can be parallel to the field, while the other Rashba contribution as well as the Ising SOC are perpendicular to it. Calculating the susceptibility for the normal state gives us both Pauli and van Vleck contributions to the susceptibility as
\begin{equation}
 \left( \left. \frac{\partial \xi_{\zeta,h}}{\partial h} \right|_{h=0} \right)^2 =  f(\theta), \qquad  \left. \frac{\partial^2 \xi_{\zeta,h}}{\partial h^2} \right|_{h=0} = \left( 1 - f(\theta) \right) \frac{\zeta}{|\lambda (\theta)|}
\end{equation} 
where the functions
\begin{equation}
    f(\theta) = \left(\frac{\lambda_\parallel (\theta) }{\lambda(\theta)} \right)^2, \qquad 1 - f(\theta) = \left(\frac{\lambda_\perp (\theta) }{\lambda(\theta)} \right)^2
\end{equation}
For singlet superconductivity the susceptibility can never reach beyond the value of the normal state van Vleck-susceptibility\cite{Maruyama2012, Sigrist2014, Skurativska2021}, as the remaining Pauli susceptibility originates from the Fermi surface. We can assume that the effect of the perpendicular SOC can be confide to the angular dependent factor $f(\theta)$ as:
\begin{equation}
   \left( \left. \frac{\partial E_{\zeta,h}}{\partial h} \right|_{h=0} \right)^2 =  f(\theta), \qquad  \left. \frac{\partial^2 E_{\zeta,h}}{\partial h^2} \right|_{h=0} \approx \left( 1 - f(\theta) \right) \frac{\zeta}{ |\lambda (\theta)|} \left(  \frac{\xi_\zeta}{E_\zeta} + \frac{\Delta^2}{\xi_k E_\zeta}\right)
\end{equation} 
As a result, the susceptibility difference for a pocket has the form 
\begin{equation}\label{eq:dchiGenR}
    \delta \chi (\theta) = f(\theta)  \left( \chi_P - \chi_{\text{sg}} \right) + \left( 1 - f(\theta)  \right) \delta \chi_2 ( \theta )
\end{equation}
We can observe that for the $\Gamma$-pocket:
\begin{equation}\label{eq:fInt}
   \int d \theta f_\Gamma( \theta ) = \int d \theta \frac{\lambda_R^2 \cos ^2 \left( \theta \right)}{\lambda_R^2 + \lambda_I^2 \cos ^2 \left( 3 \theta \right)}
\end{equation}
and for the $K$-pockets:
\begin{equation}
   \int d \theta f_K( \theta  ) = \int d \theta \frac{\lambda_R^2 \cos ^2 \left( \theta \right)}{\lambda_R^2 + \lambda_I^2 } =  \pi \left( \frac{\lambda_R}{\lambda } \right)^2
\end{equation}
where $\lambda= \sqrt{ \lambda_R^2 + \lambda_I^2 }$. The interband terms for superconducting susceptibility now depend on the full SOC $\lambda(\theta)$:
\begin{align}
   \chi_{s,0}= \sum_\zeta \int d \xi_k d \theta \left( 1 - f(\theta) \right) \frac{\zeta}{ |\lambda (\theta)|} \left( 1- \frac{\xi_\zeta}{E_\zeta} -  \frac{\Delta^2}{\xi_k E_\zeta}\right) =\\
    = -\frac{N(0)}{2 \pi}  \int d \theta \frac{ 1 - f(\theta) }{|\lambda (\theta)|} \left( - \sqrt{(\epsilon + |\lambda (\theta)| )^2 + |\Delta|^2} + \sqrt{(\epsilon - |\lambda (\theta)| )^2 + |\Delta|^2} \right)\\ \notag
     -\frac{N(0)}{2 \pi} \int d \theta \frac{\left( 1 - f(\theta) \right) \Delta^2}{|\lambda (\theta)|  \sqrt{\Delta^2 + |\lambda (\theta)| ^2}} \left[ \text{arctanh} \left( \frac{|\lambda (\theta)|^2 + |\lambda (\theta)|  \epsilon + \Delta^2 }{\sqrt{\Delta^2 + |\lambda (\theta)| ^2} \sqrt{\left( \epsilon + |\lambda (\theta)|\right)^2 + \Delta^2 }}\right)  \right.\\ \notag
    \left. -  \text{arctanh} \left( \frac{|\lambda (\theta)|^2 - |\lambda (\theta)|  \epsilon + \Delta^2 }{\sqrt{\Delta^2 + |\lambda (\theta)| ^2} \sqrt{\left( \epsilon - |\lambda (\theta)|\right)^2 + \Delta^2 }}\right) \right]
\end{align}
where if $\epsilon \rightarrow \infty$:
\begin{equation}
   \chi_{s,0} \approx \int d \theta  \left( 1 - f(\theta) \right) \chi_P  -\frac{2 N(0)}{2 \pi} \int d \theta \frac{\left( 1 - f(\theta) \right) \Delta^2}{|\lambda (\theta)|  \sqrt{\Delta^2 + |\lambda (\theta)| ^2}} \text{arctanh} \left( \frac{|\lambda (\theta)|}{\sqrt{\Delta^2 + |\lambda (\theta)| ^2} }\right).
\end{equation}
The second interband term is
\begin{equation}
   \chi_{s,T}= \sum_\zeta \int d \xi_k d \theta \left( 1 - f(\theta) \right) \frac{\zeta}{ |\lambda (\theta)|} \left(  \frac{\xi_\zeta}{E_\zeta} +  \frac{\Delta^2}{\xi_k E_\zeta}\right) F(E_\zeta).
\end{equation}
\begin{figure}
    \centering
    \includegraphics[width=0.32\linewidth]{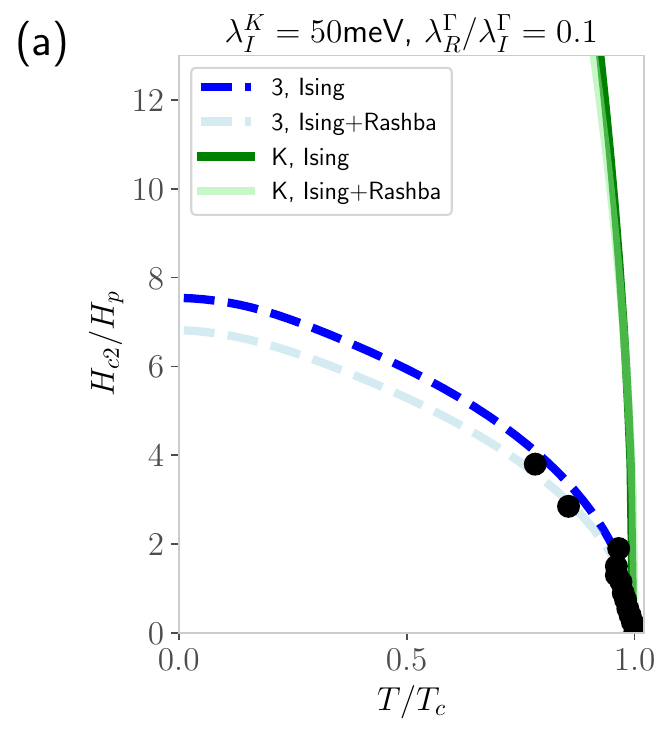}
    \includegraphics[width=0.41\linewidth]{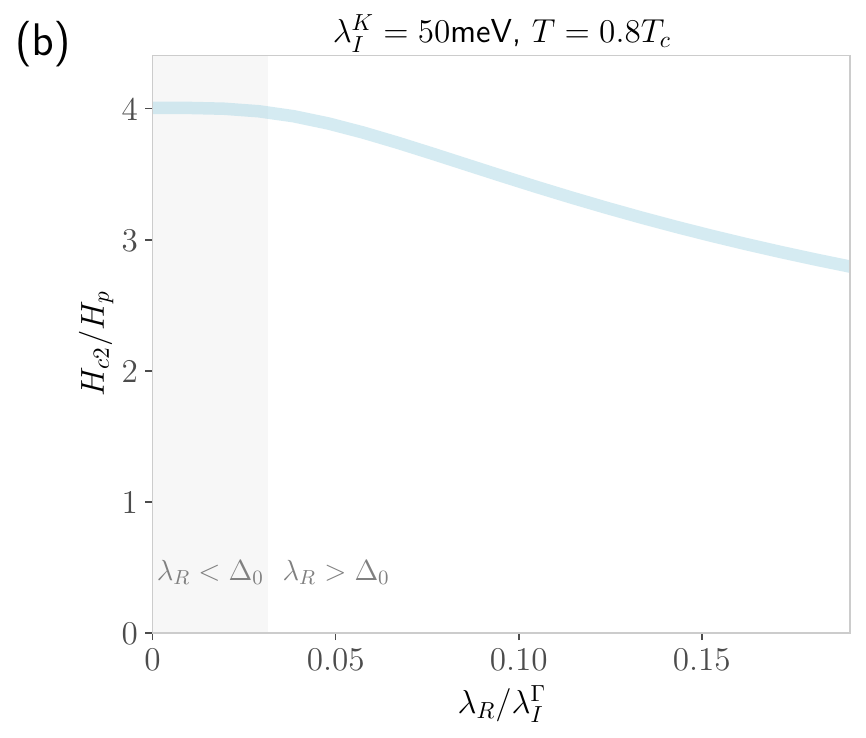}
    \caption{\label{fig:Rashba_suscp} Ising and Rashba SOC for a uniform singlet order. (a) The critical field for a in-plane field is compared when a Rashba SOC was included both for all 3 pockets and for a model with only $K$-pockets. The value of the Rashba SOC is fixed to $\lambda_R^\Gamma = 2 \lambda_R^K = 0.1 \lambda_I^\Gamma $. (b) The critical field for the 3 pocket model $H_{c2} (T=0.8T_c)$, evaluated via numerical integration of the susceptibility difference, is suppressed by the Rashba SOC in a roughly linear way for a small $\lambda_R / \lambda_I$. }
\end{figure}
For $K$-pockets the modification of the susceptibility difference from a Rashba SOC can be easily seen to be of order $(\lambda_R / \lambda_I)^2$:
\begin{equation}\label{eq:dchiKR}
    \delta \chi^K_R (T)= \frac{1}{2}\left( \frac{\lambda_R^K}{\lambda^K } \right)^2 \left( \chi_P - \chi_{\text{sg}} (T) \right) + \left( 1 - \frac{1}{2}\left( \frac{\lambda_R^K}{\lambda^K } \right)^2 \right) \delta \chi^K_2 ( T )
\end{equation}
where $\delta \chi^K_2 ( T )$ is the form of the susceptibility difference for only Ising SOC but where the total SOC is $\lambda^K$, such that $\delta \chi^K_2 ( T ) \propto \left( \frac{\Delta_0^K}{\lambda^K} \right)^2$. For the $\Gamma$-pocket we can approximate the full susceptibility difference as:
\begin{align}\label{eq:dchiGR}
    \delta \chi^\Gamma_R (T) \approx \int d \theta \left( f_\Gamma (\theta)+ \left( 1 - f_\Gamma(\theta) \right)  \alpha_{0} (\theta) \right)  \left( \chi_P - \chi_{\text{sg}}(T) \right) 
\end{align}
with $\alpha_{0} (\theta) \propto \frac{\Delta_0^\Gamma}{\lambda_I^\Gamma} $, calculated as a peak around the nodal points of the Ising SOC.

To get a scaling of $\delta \chi^\Gamma ( T )$  with $\lambda_R$, we consider both limits $\lambda_R \ll \Delta_0$ and $\lambda_R \gg \Delta_0$ when expanding $\alpha_{0} (\theta)$ close to a node with a Lorentzian (Eq.~\eqref{eq:LorGen}):
\begin{equation}
   \alpha_{0}(\theta - \theta_0)= \alpha L(\theta - \theta_0, \beta) = \frac{\alpha}{\beta \pi} - \frac{\alpha}{\beta^3 \pi} (\theta - \theta_0)^2+ \dots
\end{equation}
For the susceptibility $\chi_{s,0}$ at $\epsilon \rightarrow \infty$, the constant $\alpha$ expanded in small $\lambda_R$ is:
\begin{align}
    \alpha \approx \frac{\pi}{\sqrt{3}} \frac{\Delta_0}{\lambda_I} \left( 1 - \frac{\lambda_R^2}{5 \Delta_0^2}\right)
\end{align}
where the correction from the Rashba SOC is of order $\frac{\lambda_R^2}{\Delta_0^2}$. However, it is also possible that Rashba is large compared to the superconducting order. If we first expand $\chi_{s,0}$ for a small $\Delta$ and then expand that term close to a node $\theta_0$, we instead obtain:
\begin{align}
    \alpha \approx \frac{\pi}{\sqrt{3}} \frac{\Delta_0}{\lambda_I} \left( \frac{\Delta_0}{\sqrt{6} \lambda_R}  \text{ln} \frac{4 \lambda_R^2}{\Delta_0^2} \sqrt{\frac{\text{ln} \frac{\Delta_0^2}{4 \lambda_R^2}}{1 + \text{ln} \frac{\Delta_0^2}{4 \lambda_R^2}}} \right)
\end{align}
The total value $\delta \chi_R (T)$ therefore has the additional dependence, beyond $\frac{\Delta}{\lambda_I}$, on the two ratios $\frac{\lambda_R}{\lambda_I}$ and $\frac{\Delta}{\lambda_R}$, at each pocket. To determine the correction to the critical field at small $\lambda_R$, using the Eq.s~\eqref{eq:dchiKR} \& \eqref{eq:dchiGR}
we approximate $\int d \theta f_\Gamma(\theta) \approx \lambda_R / (2 \lambda)$ and get:
\begin{equation}
    \frac{H_{c2}^K (T)}{H_p} \approx \frac{H_{c2}^{K, I} (T)}{H_p} \left( 1 + \frac{1}{4} \left( \frac{\lambda_R^K}{\lambda^K } \right)^2 - \frac{1}{2} \left( \frac{\lambda_R^K}{\Delta^K} \right)^2 \frac{1 - Y(T)}{\ln \left( \left( \frac{\lambda^K}{\Delta^K} \right)^2  -1\right)}\right)
\end{equation}
\begin{equation}
    \frac{H_{c2}^\Gamma (T)}{H_p} \approx \frac{H_{c2}^{\Gamma, I} (T)}{H_p} \left( 1 + \frac{1}{4} \left| \frac{\lambda_R^\Gamma }{\lambda^\Gamma  } \right| - \frac{1}{4} \left| \frac{\lambda_R^\Gamma }{\Delta_0^\Gamma } \right| \left( 1 - Y(T) \right) \right) 
\end{equation}
The effect of the Rashba SOC on the critical field is two types of corrections: an increase proportional to $\frac{\lambda_R}{\lambda }$, from the increased splitting between bands, and a decrease proportional to $\frac{\lambda_R}{\Delta}$, from the now present SOC parallel to the magnetic field. Since $\lambda > \Delta$, this results in a overall suppression of the critical field. As shown in Fig.~\ref{fig:Rashba_suscp}(a) a Rashba SOC of the size $\lambda_R^\Gamma = 0.1 \lambda_I^\Gamma \approx 1.7$meV suppresses the critical field for monolayer NbSe$_2$ when all three pockets are considered. In Fig.~\ref{fig:Rashba_suscp}(b) the critical field is shown to be suppressed by the Rashba SOC both when $\lambda_R > \Delta_0$ and $\lambda_R < \Delta_0$.

\section{Role of the Q-pockets in MoS$_2$}\label{sec:Qpockets}
\begin{figure}
    \centering
    \includegraphics[width=0.35\linewidth]{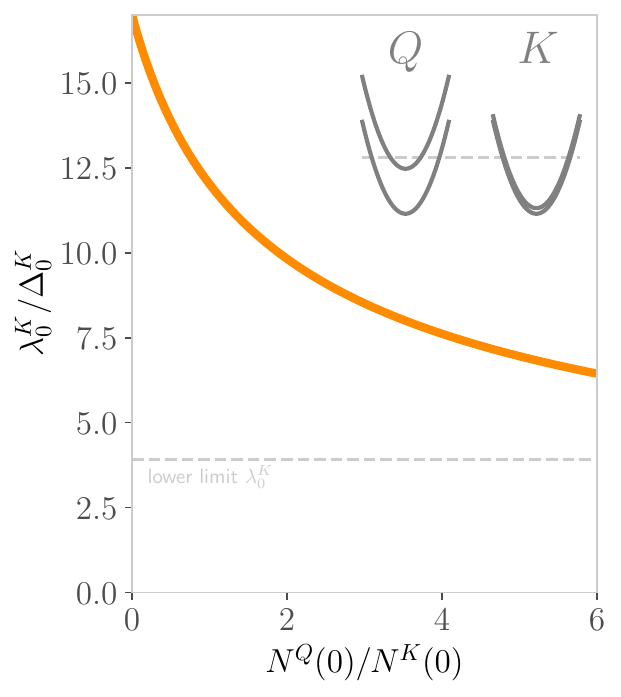}
    \caption{\label{fig:KQcond}For $K$- and $Q$-pockets present in MoS$_2$ the     upper critical field, Eq.~\eqref{eq:KQcrit}, is set to be equal to that of only $K$-pockets $\lambda_0^K /\Delta_0^K=17$\cite{Lu2015}. Even for a large density of states $ N^Q (0)$ a realistic value of the band splitting at the FS, $\lambda_0^K$, is possible.}  
\end{figure}
The full superconducting dome for gated MoS$_2$ has been revealed to not only have the $K$-pockets but additional pockets around the $Q$-points\cite{Marini2023}, located halfway between $\Gamma$ and $K$. The density of states belonging to each type of pocket is dependent on the gating-induced doping, where several works have predicted that the $Q$-pocket is not only present but crucial for the superconducting pairing mechanism\cite{Costanzo2016,Marini2023}. Of note is the constant spin splitting of both types of pockets where $\lambda_0^Q \approx 30 \lambda_0^K $, with a splitting of 1.5meV at the $K$-point\cite{Marini2023}. However, for the samples in which the critical field has been measured the exact filling of each pocket is unknown.

The presence of the two types of pockets is an extension of the idea previously exemplified when the $\Gamma$-pocket is present: segments of the FS with the lowest SOC will determine the critical field. The critical field is determined by the total susceptibility at the FS with $ \Omega^{\text{tot}}_0(T) = \Omega^{Q}_0(T) + \Omega^{K}_0(T)$ and $\delta \chi^{\text{tot}}(T) = \delta \chi^{Q}(T) + \delta \chi^{K}(T)$. Both pockets have a constant Ising SOC, where the one with the much lower splitting is dominating the signal. Since it is predicted that $\Delta^K \approx \Delta^Q$\cite{Marini2023} the susceptibility difference originating from the $Q$-pockets becomes negligible:
\begin{align}\label{eq:KQcrit}
    \frac{H_{c2} (T)}{H_p} \propto \sqrt{\frac{N^K (0) + N^Q (0) (\Delta_0^Q / \Delta_0^K)^2}{N^K (0) \delta \chi_2^K (T)+ N^Q (0) \delta \chi_2^Q (T) }} \approx \sqrt{1 + \frac{N^Q  (0)}{N^K  (0)}}  \frac{ \lambda_0^K }{\Delta_0^K}
\end{align}
as $\delta \chi_2^j (T) \propto \left( \frac{\Delta^j}{ \lambda_0^j }\right)^2$ and $\delta \chi_2^K (T) \approx 30^2 \delta \chi_2^Q (T) $.

In Fig.~\ref{fig:KQcond} the DoS of the $K$- and $Q$- pockets are varied and the value of $\lambda_0^K /\Delta_0^K$ is shown for which the critical field has the same value as for the single pocket approximation with $\lambda_0^K =6.2$meV\cite{Lu2015} ($\lambda_0^K /\Delta_0^K=17$). The splitting $\lambda_0^K $ at the FS for the $K$-pockets will vary with doping and can be as small as the value at the $K$-point. Even for a minimal presence of the $K$-pockets at the FS they determine the upper critical field.

A final point for the critical field measurements in gated MoS$_2$ is the validity of considering the system as a monolayer\cite{Costanzo2016}. For the few-layer samples a suppression of the critical field from interlayer hopping could be present. Such an effect would, similarly to the NbSe$_2$ case, affect both types of pockets.

\section{Condensation energy}\label{sec:condEn}
The condensation energy is calculated as the difference in the thermodynamic potential between the normal $\Omega_N (T,h)$ and superconducting state $\Omega_S (T,h)$, at $h=0$ and a non-zero temperature $T$\cite{Ortega2020}:
\begin{align}\label{eq:PotFull}
  \Omega_0 (T) = \Omega_S (T,0) - \Omega_N (T,0)= 2 k_{\text{B}} T \sum_{\zeta, \bk} \ln \left( \frac{1 + e^{- | \xi_{\bk,\zeta} |/ k_{\text{B}} T} }{1 + e^{- E_{\bk,\zeta}/ k_{\text{B}} T}} \right)\\ \notag
    + \sum_{\zeta,\bk} \left[ | \xi_{\bk,\zeta} | - E_{\bk,\zeta} + \frac{| \Delta_{\bk,\zeta}|^2}{2 E_{\bk,\zeta}} \tanh \frac{E_{\bk,\zeta}}{2  k_{\text{B}} T }\right],
\end{align}
for two bands labeled by $\zeta=\pm$ and Bogoliubov-de Gennes (BdG) bands are denoted by $E_{\bk,\zeta}$. In this section we demonstrate that the spin-orbit coupling $\lambda (\bm{k})$ has a negligible effect on the condensation energy. The calculation is performed for a constant singlet pairing $\Delta_{\bk} = \Delta$, simplifying
\begin{equation}
     \sum_{\bk, \zeta} \frac{| \Delta_{\bk,\zeta}|^2}{2 E_{\bk,\zeta}} \tanh \frac{E_{\bk,\zeta}}{2  k_{\text{B}} T } = \frac{ \Delta(T)^2}{2}
\end{equation}
We first establish that without SOC $\Omega_{0,\lambda=0}=\Omega_{0,0, \lambda=0} + \Omega_{0,T, \lambda=0}$ where the bands are integrated in a range $\pm \epsilon$:
\begin{equation}
      \Omega_{0,0, \lambda=0} - N(0) \Delta(0,T)^2 =  2\sum_{\bk} \left[  | \xi_{\bk} | - E_{\bk} \right]
     = - N(0) \Delta^2 \left[ \frac{\epsilon}{\Delta} \sqrt{1 + \left( \frac{\epsilon}{\Delta} \right)^2} - \left( \frac{\epsilon}{\Delta} \right)^2 +  \text{arcsinh} \left( \frac{\epsilon}{\Delta}\right) \right]
\end{equation}
Expanded in small $\Delta / \epsilon$ the remaining terms are
\begin{equation}
      \Omega_{0,0, \lambda=0} \approx - N(0) \frac{1}{2} \Delta^2 \left[ 1 +2 \text{ln} \left( \frac{\epsilon}{\Delta}\right) \right]
\end{equation}
\begin{equation}
      \Omega_{0,T, \lambda=0}  = -2 N(0) k_B T \Delta \int_{-\epsilon/\Delta}^{\epsilon/\Delta} ds \text{ln} \left[ 1 + e^{-\Delta \sqrt{1 + s^2}/k_B T }\right] + \frac{\pi^2}{3} N(0) \left( k_B T \right)^2
\end{equation}
Close to $T=T_c$ the following integral can be solved approximately as
\begin{equation}
    I(\delta) = \frac{1}{\delta} \int_0^{\infty} ds~ \text{ln} \left[ 1 + e^{-\delta \sqrt{1 + s^2} }\right] 
    = \frac{\pi^2}{12 \delta^2} + \frac{1}{4} \text{ln} \frac{2 \delta}{\pi} + \frac{1}{4} (C - \text{ln}2 - \frac{1}{2}) - \frac{7 \zeta (3)}{64 \pi^2} \delta^2 + \mathcal{O} (\delta^4)
\end{equation}
where $\delta = \frac{\Delta}{k_B T}$, which is valid with the approximation $\epsilon/ \Delta \rightarrow \infty$. Combining the terms, we see that close to $T_c$:
\begin{equation}
     \Omega_{0, \lambda=0} =  \Omega_{0,0, \lambda=0} + \Omega_{0,T, \lambda=0}  \approx N(0) \text{ln} \left( \frac{T}{T_c}\right) \Delta^2 + \frac{7 \zeta (3)}{16 \pi^2} \frac{\Delta^4}{\left( k_B T \right)^2}
\end{equation}
If we now turn to a finite SOC $\lambda_0$, there are two bands shifted by $\pm \lambda_0$. By substitution of variables the integration range in $\Omega_{0,T}$ can be shifted to get terms on the form:
\begin{align}
      \Omega_{0,T}  = - N(0) k_B T \Delta \left(\int_{\frac{-\epsilon+ \lambda_0}{\Delta}}^{\frac{\epsilon+ \lambda_0}{\Delta}} ds ~\text{ln} \left[ 1 + e^{-\Delta \sqrt{1 + s^2}/k_B T }\right] + \int_{\frac{-\epsilon- \lambda_0}{\Delta}}^{\frac{\epsilon- \lambda_0}{\Delta}} ds ~\text{ln} \left[ 1 + e^{-\Delta \sqrt{1 + s^2}/k_B T }\right]\right) + \frac{\pi^2}{3} N(0) \left( k_B T \right)^2\\ \notag
     = - N(0) k_B T \Delta \left(\int_{\frac{-\epsilon+ \lambda_0}{\Delta}}^{\frac{\epsilon- \lambda_0}{\Delta}} ds~ \text{ln} \left[ 1 + e^{-\Delta \sqrt{1 + s^2}/k_B T }\right] + \int_{\frac{-\epsilon- \lambda_0}{\Delta}}^{\frac{\epsilon+ \lambda_0}{\Delta}} ds~ \text{ln} \left[ 1 + e^{-\Delta \sqrt{1 + s^2}/k_B T }\right]\right) + \frac{\pi^2}{3} N(0) \left( k_B T \right)^2\\
\end{align}
If we now make an assumption about the integration range $\epsilon$, that both $\frac{\epsilon+ \lambda_0}{\Delta} \rightarrow \infty$ and $\frac{\epsilon- \lambda_0}{\Delta} \rightarrow \infty$, this is exactly the same result as for $ \Omega_{0,T, \lambda=0}$. For the other term
\begin{equation}
     \Omega_{0,0} - \frac{ \Delta(T)^2}{ v}=  \sum_{k, \zeta} \left(  | \xi_{\bk,\zeta} | - E_{\bk,\zeta} \right)
     =  N(0) \frac{ \Delta (T)^2 }{2} \left(1 + 2~\text{arcsinh} \left( \frac{\epsilon + \lambda_0}{\Delta}\right) + 2 ~\text{arcsinh} \left( \frac{\epsilon - \lambda_0}{\Delta}\right) \right)
\end{equation}
and expanding $ \Omega_{0,0}$ for $\frac{\epsilon}{\Delta} \gg \frac{\lambda_0}{\Delta}$ results in small corrections
\begin{equation}
    \text{arcsinh} \left( \frac{\epsilon + \lambda_0}{\Delta}\right) + \text{arcsinh} \left( \frac{\epsilon - \lambda_0}{\Delta}\right) \approx 2 ~\text{arcsinh} \left( \frac{\epsilon}{\Delta}\right) -  N(0) \frac{\epsilon \lambda_0^2}{\left( \epsilon^2 + \Delta^2\right)^{3/2}}
\end{equation}
Hence, the condensation energy at any given temperature is:
\begin{equation}
     \Omega_{0} \approx \Omega_{0, \lambda=0} - \frac{\Delta^2 \lambda_0^2}{ \epsilon^2} \approx \Omega_{0, \lambda=0} 
\end{equation}
For a finite $\epsilon$ a numerical calculation of the condensation energy gives us no discernible differences between bands with or without SOC. As the condensation energy does not scale with SOC, the same expression can be used for pockets at the FS with either constant or nodal SOC.

\bibliography{HC_ising_G_refs}